\documentstyle[12pt]{article}
\input latexsym.sty
\begin{document}

\hoffset = -1truecm
\voffset = -2truecm

\title{\bf
The Schwinger Model on the Torus
}

\author{
{\bf
S.Azakov}\thanks{Permanent Address: Institute of Physics, Azerbaijan
Academy of Sciences, Baku, Azerbaijan}\\ International Center for Theoretical
Physics, Trieste 34100, {\bf Italy}\\ and\\
Institute for Advanced Studies in Basic Sciences, Zanjan, {\bf
Iran}\thanks{Present
Address.} }

\date{6 August 1996}
\newpage

\maketitle

\newcommand{\ren}{\renewcommand{\theequation}{1.\arabic{equation}}}
\newcommand{\new}[1]{\renewcommand{\theequation}{1.\arabic{equation}#1}}
\newcommand{\add}{\addtocounter{equation}{-1}}

%\vspace{3cm}
\begin{abstract}
The classical and quantum aspects of the Schwinger model on the torus are
considered. First we find explicitly all zero modes of the Dirac operator in the
topological sectors with nontrivial Chern index and its spectrum. In the
second part
we determine the regularized effective action and discuss the propagators
related
to it.

Finally we calculate the gauge invariant averages of the fermion bilinears and
correlation functions of currents and densities. We show that in the
infinite volume
limit the well-known result for the chiral condensate can be obtained and the
clustering property can be established.
\end{abstract}
\vspace{3cm}

Manuscript pages 40.

\newpage

\newpage

Proposed running head: "The Schwinger model on the Torus"

Mailing Address:

S.Azakov

Institute for Advanced Studies in Basic Sciences,

Gaveh Zang, P.O.Box 45159-195

Zanjan, IRAN

Telephone:(+98)(241)449021 (Iran),

Fax:	(+98)(241)449023 (Iran)

e-mail: azakov@rose.ipm.ac.ir

\vspace{2cm}

{\bf Introduction}

The Schwinger Model (SM) \cite{Schwi} or quantum electrodynamics with massless
fermions in two-dimensional space-time is one of the exactly solvable models of
quantum field theory (QFT). This was shown some time ago by using operator
methods \cite{Low} and the path integral approach \cite{NS}%
\footnote{The
discussion of the SM are so numerous that it is not possible to refer to
all of them.}.
The SM usually serves for the illustration of such phenomena in particle
physics as:
spontaneous breakdown of local gauge symmetry through axial anomaly \cite{KJ},
mass generation, charge screening \cite{CKS} and quark confinement, vacuum
structure and the realization of gauge transformations \cite{NS},\cite{RW}.

Most of the mathematical problems of QFT, related to these phenomena can be
treated more exactly and rigorously in a compactified version of Euclidean
space-time, where the spectrum of the Dirac operator becomes exact, and a
precise
definition of topological sectors together with corresponding zero modes can be
given%
\footnote{Notice that technically compactification allows to avoid the infrared
divergences, which sometimes plague the analysis of two-dimensional gauge
theories.}. Such compactification in the SM on the manifold without boundary was
considered for the first time by C.~Jayewardena \cite{Camil}, who used
two-dimensional sphere as a
compact Euclidean space-time.

In this work we present the detailed calculations concerning classical and
quantum
aspects of the SM on the torus.

Compactification on the torus ($\cal T$) is much better than the
compactification on
the sphere because it allows to find finite temperature (size) effects and is
appropriate to the systematic analysis of the lattice approximation on which the
numerical calculations are performed (on the lattice, usually periodic (torus)
boundary conditions are considered). Moreover, from the results obtained
for this
model one can extract the information concerning the SM on a cylinder,
which also
has attracted  attention recently \cite{Man}.

We show that the torus compactification on the SM, as in the case of sphere
compactification, does not destroy the solvability of the model.

The SM on the symmetric torus (where the lengths of both circumferences are
equal) has been considered in \cite{Aza} and the SM with Dirac-K\"ahler fermions
(the geometric SM, which is equivalent to the SM with two flavors) on the
torus was
investigated in \cite{JO1} and \cite{JA}. I.Sachs and A.Wipf \cite{SW}
discussed the
role of the zero modes in the SM on the torus, derived some relevant Green's
functions and found the analytic form of the chiral condensate.

Finite temperature SM has been considered in \cite{Kao}.

The paper is organized as follows.

The first part contains the definition of the model and the discussion of some
peculiarities due to the fact that it is defined on the compact space-time.
We find the
general expression for the normalized zero modes of the Dirac operator in the
sector with any topological charge. Then we calculate the spectrum of this
operator
and the corresponding eigenfunctions.

In the second part we discuss the general path integral formula which can
be used
for the calculation of the quantum mechanical vacuum expectation values of
observables in the case where zero modes are present. The rest of this part is
devoted to the calculation of the regularized effective action and the
propagators
related to it.

In the third part we consider the objects of physical significance, i.e.
the expectation
values of various operators.

We calculate separately the contributions to them from different
topological sectors.
Taking the infinite volume limit we obtain the well-known result for the chiral
condensate and establish the clustering property.

In the appendices the details of the derivations of some important formulas
used in the
main text are presented.

\section{Classical theory}
\subsection{The action}

\ren
\begin{equation}
S=\int\limits_{{\cal T}=S_1\times S'_1} d^2x \left\{\frac{1}{2}F_{12}^2(x)
+\bar{\psi}(x)
\gamma_{\mu}(\partial_{\mu}-ieA_{\mu})\psi(x)\right\},
\label{a11}\end{equation}
where $0 \leq x_{\mu} \leq L_{\mu},\;\mu=1,2 $, $L_1$ and $L_2$ are lengths
of the
large and small
circumferences of the torus, respectively, and the field strength
$F_{12}(x)=\partial_1 A_2(x)-\partial_2 A_1(x)$.

Our $\gamma$ -matrices
satisfy:$\{\gamma_{\mu},\gamma_{\nu}\}=2\delta_{\mu\nu},
\gamma_1\gamma_2=i\gamma_5,\gamma_{\mu}^{\dagger}=\gamma_{\mu}$, which
implies in two dimensions $\gamma_{\mu}\gamma_5=-i\epsilon_{\mu\nu}
\gamma_{\nu}$, where $\epsilon_{12}=-\epsilon_{21}=1$.

The geometry of fields on the torus requires certain periodicity conditions:

\new{a}
\begin{equation}
\psi(x+L_\nu\hat{\nu})=\Lambda_\nu(x)\psi(x),\label{asd} \end{equation} \add
\new{b}
\begin{equation}
\bar{\psi}(x+L_\nu\hat{\nu})=\bar{\psi}(x)\Lambda_\nu^{-1}(x),
\label{bsd}\end{equation}
\ren
\begin{equation}
A_\mu(x+L_\nu\hat\nu)=A_\mu(x)-\frac{i}{e}\Lambda_\nu^{-
1}(x)\partial_{\mu}\Lambda_{\nu}(x), \label{a13}\end{equation} where
$\hat{\nu}$ is
a unit vector in $\nu^{th}$ direction. The transition functions
$\Lambda_{\nu}(x)$ satisfy the {\it cocycle condition} \begin{equation}
\Lambda_\mu(x)\Lambda_\nu(x+L_\mu\hat{\mu})=\Lambda_\nu(x)\Lambda_\mu(x+
L_\nu\hat{\nu}). \label{a14}\end{equation}

It is well-known \cite{Gil} that under these requirements the gauge field
configurations fall into classes ${\cal CH}^{(k)}$ (topological sectors,
{\it Chern
classes}) characterized by the {\it Pontriyagin (Chern) index} (topological
charge,
topological quantum number, winding number) \begin{equation}
k=\frac{e}{2\pi}\int_{\cal T}F_{12}d^2x, \;\;\;\;k=0,\pm1,\pm2,\cdots
\label{a15}\end{equation}

As a special representative we choose a field with a constant field strength:
\begin{equation}
F_{\mu\nu}=B\epsilon_{\mu\nu}=\frac{2\pi k}{eL_1L_2}\epsilon_{\mu\nu}.
\label{a16}\end{equation}
In the {\it Lorentz} gauge a corresponding potential is \begin{equation}
C_\mu^{(k)}(x)=-\frac{B}{2}\epsilon_{\mu\nu}x_\nu=-\frac{\pi
k}{eL_1L_2}\epsilon_{\mu\nu}x_\nu. \label{a17}\end{equation}

In the {\it axial} gauge $\partial_1A_\mu(x)=0$ we may choose the
representative

$$
\tilde{C}^{(k)}_\mu(x)=\left\{
\begin{array}{cl}
-\frac{2\pi k}{eL_1L_2}x_2, &\mbox{ for $\mu=1$} \\ 0, & \mbox{ for $\mu=2$}.
\end{array}\right.
$$

In this case
$$ \tilde{F}^{(k)}_{12}=\partial_1\tilde{C}_2^{(k)}(x)
-\partial_2\tilde{C}^{(k)}_1(x)=\frac{2\pi}{e}\frac{k}{L_1L_2}$$ and we see
that again
$$
\frac{e}{2\pi}\int_{\cal T}\tilde{F}_{12}d^2x=k.$$ The periodicity condition is
$$\tilde{C}^{(k)}_\mu(x+L_\nu\hat{\nu})=\tilde{C}_\mu^{(k)}(x)-
\frac{i}{e}\Lambda_\nu^{-1}(x)\partial_\mu\Lambda_\nu(x),$$ where $$
\Lambda_\nu(x)=\left\{
\begin{array}{cl}
1,& \mbox{ for $\nu=1$} \\
e^{-2\pi ik\frac{x_1}{L_1}},&\mbox{ for $\nu=2$}. \end{array}\right. $$

The relation to the Lorentz gauge is

$$ C_\mu^{(k)}(x)=\tilde{C}^{(k)}_\mu(x)-\frac{i}{e}E^{-1}(x) \partial_\mu
E(x),$$
where
$$E(x)=e^{i\pi k\frac{x_1x_2}{L_1L_2}}.$$

In what follows we will work in the Lorentz gauge.

In the {\it topological sector} ${\cal A}^{(k)}$ a general gauge potential
has the form
\begin{equation}
A_\mu^{(k)}(x)=A_\mu^{(0)}+C_\mu^{(k)}(x), \label{a18}\end{equation}
$A_\mu^{(0)}(x)$ is a single valued `continuous' function on ${\cal T}$. Thus we
may apply the Hodge decomposition theorem \cite{Gil} \begin{equation}
A_\mu^{(0)}(x)=\partial_\mu a(x)+t_\mu+\epsilon_{\mu\nu}\partial_\nu b(x),
\label{a19}\end{equation}
where $\partial_\mu a(x)$ is a `pure gauge', $t_\mu$ is a (constant) toron field
restricted to $ 0\leq t_\mu < 2\pi/eL_\mu $, $
\epsilon_{\mu\nu}\partial_\nu b(x)$ is a
curl and $a(x)$ and $b(x)$ are continuous on ${\cal T}$ and orthogonal to the
constant, $\int_{\cal T} a(x)d^2x=\int_{\cal T} b(x)d^2x=0.$

The toron field is a gauge invariant up to {\it large gauge
transformations}, when
$\Lambda(x)=\exp\left\{2\pi i\left(m_1\frac{x_1}{L_1}+m_2\frac{x_2}{L_2}
\right)\right\}$, and $m_\mu$ is integer: $t_\mu{\mathop{\rightarrow}
\limits^\Lambda} t_\mu+\frac{2\pi}{eL_\mu}m_\mu$, in this case
$a(x){\mathop{\rightarrow}\limits^\Lambda}a(x)$ and
$b(x){\mathop{\rightarrow}\limits^\Lambda}b(x)$. Under {\it small gauge
transformations} $\Lambda(x)=e^{i\lambda(x)}$, $a(x){\mathop{\rightarrow}
\limits^\Lambda}a(x)+\frac{1}{e}\lambda(x)$, $b(x){\mathop{\rightarrow}
\limits^\Lambda}b(x)$, $t_\mu{\mathop{\rightarrow}\limits^\Lambda}t_\mu$.

\subsection{Dirac equation (DE)}

\begin{equation}
D\chi(x)\equiv\gamma_\mu(\partial_\mu-ieA_\mu)\chi(x)=0,
\label{a110}\end{equation}
where
\[\chi(x)={\chi_1(x) \choose \chi_2(x)}\] is a two-component complex spinor.

Formally the solution of DE (\ref{a110}) with the potential (\ref{a18}),
(\ref{a19}) is
\begin{equation}
\chi(x)=e^{ie[a(x)+t_{\mu} x_{\mu}-i\gamma_5(b(x)-\frac{\pi k}{2eL_1L_2}
x^2)]}\;^0\chi(x)
\label{a111}\end{equation}
and $^0\chi(x)$ satisfies the free DE
\begin{equation}
\gamma_\mu\partial_\mu^0\chi(x)=0.
\label{a112}\end{equation}
The main problem is to find such $^0\chi(x)'s$, which satisfy the periodicity
condition (1.2) for $\chi(x)$.

In the future we will consider also the operator \begin{equation} D_0=\left.
D\right|_{a=b=0}=\gamma_\mu(\partial_\mu-ie(t_\mu+C_\mu^{(k)}(x))).
\label{a113}\end{equation}
Then

\[\hat{\chi}_i(x)=e^{e\gamma_5b(x)+iea(x)}\chi_i(x),\] where
$\hat{\chi}_i(\chi_i)$ is a
zero mode of the $D(D_0)$ operator.

\subsection{General remarks}

The operator $iD$ is the Hermitian operator in the space of two - component
spinors, with a scalar product
\[ (\psi,\chi)=\int_{\cal T} d^2x\bar{\psi}(x)\chi(x). \] It is an
{\bf elliptic operator} on the
compact manifold ${\cal T}$ and its spectrum is {\bf discrete}.
$\{\psi_\nu\}(\nu=1,2,\cdots)$ is a set of independent eigenfunctions of
the $iD$ with
positive eigenvalues $E_\nu$. Since $iD$ anticommutes with $\gamma_5$,
$\{\gamma _5\psi_\nu\}(\nu=1,2,\cdots)$ is a set of independent eigenfunctions of
the $iD$ with
eigenvalues $-E_{\nu}$. Denoting $\psi_{-\nu}=\gamma_5\psi_{\nu}$ and $E_{-
\nu}=-E_{\nu}$ we have, that $\{\psi_{\nu}\},\nu=\pm 1,\pm 2,\cdots$ is a
complete set
of the eigenfunctions of $iD$. Again, since $\{D,\gamma_5\}=0$, we can
choose zero
modes to have a definite {\it chirality}.

{\bf The Index theorem.} Each zero mode $\chi_i(x)$ has a definite chirality
$\gamma_5\chi_i=\pm\chi_i$ and a number of the zero modes $n=n_++n_-[n_+(n_-
)$ is a number of the zero modes with a positive (negative) chirality]
satisfies the
following rule
\[n_+=k, \;\; n_-=0,\;\; \makebox{if}\: k\geq 0,\]

$$
n_+=0,\;\; n_-=|k| ,\;\; \makebox{if}\: k\leq 0. $$

In the trivial sector $(k=0)$ there are no zero modes \cite{Pat}.

\subsection{The zero modes of the $D_0$ operator in the nontrivial sector}

Let us introduce the complex (dimensionless) variables \begin{equation}
z=\frac{x_1+ix_2}{L_1},\;\;\;\;\bar{z}=\frac{x_1-ix_2}{L_1}.
\label{a114}\end{equation}
Then
\begin{equation}
\partial_1=\frac{1}{L_1}(\partial_z+\partial_{\bar{z}}),\;\;\;\;\partial_2=
\frac{i}{L_1}(\partial_z-\partial_{\bar{z}}), \label{a115} \end{equation}
and the free Dirac equation (\ref{a112}) has the following form
\begin{equation}
2\left(\begin{array}{cc}
0&\partial_z\\
\partial_{\bar{z}}&0
\end{array}
\right)\left(\begin{array}{c}
\;^0\chi_1(z,\bar{z})\\
\;^0\chi_2(z,\bar{z})
\end{array}
\right)=0 .
\label{a116}\end{equation}
{}From this equation we see that
\new{a}
\begin{equation}
\;^0\chi_1(z,\bar{z})=\;^0\chi_1(z) ,\end{equation} \add \new{b}
\begin{equation}
\;^0\chi_2(z,\bar{z})=\;^0\chi_2(\bar{z}). \label{a117}\end{equation} \ren

The vector-potential is chosen as follows \begin{equation}
A_{\mu}(x)=t_{\mu}+C_{\mu}^{(k)}(x)=t_{\mu}-\frac{\pi
k}{eL_1L_2}\varepsilon_{\mu\nu}x_{\nu}, \label{a118}\end{equation} and in
accordance with the general solution (\ref{a111}) we get \new{a}
\begin{equation}
\chi_1(z,\bar{z})=e^{i\frac{e}{2}L_1(t_-z+t_+\bar{z})-\frac{\pi}{2}k\frac{|z
|^2}
{|\tau|}}\;^0\chi_1(z),
\label{aw}
\end{equation}
\add
\new{b}
\begin{equation}
\chi_2(z,\bar{z})=e^{i\frac{e}{2}L_1(t_{-
}z+t_{+}\bar{z})+\frac{\pi}{2}k\frac{|z|^2}{|\tau|}}\;^0\chi_2(\bar{z}),
\label{bw}\end{equation}
\ren
where $\chi_1(z,\bar{z})$ and $\chi_2(z,\bar{z})$ are the components of the
spinor
\begin{equation}
\chi(z,\bar{z})={\chi_1(z,\bar{z}) \choose \chi_2(z,\bar{z})},
\label{a120}\end{equation}
which is the solution of the equation
\begin{equation}
D_0\chi(z,\bar{z})=0 ,
\label{a121}\end{equation}
\(t_{\pm}=t_1\pm it_2\).

Using the expression (\ref{a118}) and taking into account (1.2) we get the
periodicity conditions for the function $\chi(z,\bar{z})$,
\(\tau=i\frac{L_2}{L_1}\)

\new{a}
\begin{equation}
\chi(z+1,\bar{z}+1)=e^{\frac{\pi k}{2|\tau|}(z-\bar{z})}\chi(z,\bar{z}),
\label{asw}\end{equation}
\add
\new{b}
\begin{equation}
\chi(z+\tau,\bar{z}+\bar{\tau})=e^{-i\frac{\pi k}{2}(z+\bar{z})}
\chi(z,\bar{z}).\label{bsw}
\end{equation} \ren
Then from (1.19) and (1.22) it follows that $\;^0\chi_1(z)$ and
$\;^0\chi_2(\bar{z})$
have the following periodicity properties (for simplicity we put $L_1=1$):
\new{a}
\begin{equation}
\;^0\chi_1(z+1)=e^{\frac{\pi k z}{|\tau|}+\frac{\pi
k}{2|\tau|}-\frac{ie}{2}(t_++t_-)}
\;^0\chi_1(z),
\label{aq}
\end{equation}
\add
\new{b}
\begin{equation}
\;^0\chi_1(z+\tau)=e^{-i\pi k z+\frac{\pi}{2}k|\tau|-\frac{e|\tau|} {2}(t_+-t_-
)}\;^0\chi_1(z),
\label{bq}
\end{equation}
\new{a}
\begin{equation}
\;^0\chi_2(\bar{z}+1)=e^{-\frac{\pi k\bar{z}}{|\tau|}-\frac{\pi
k}{2|\tau|}-\frac{ie}
{2}(t_++t_-)}\;^0\chi_2(\bar{z}),
\label{ae}
\end{equation}
\add
\new{b}
\begin{equation}
\;^0\chi_2(\bar{z}+\bar{\tau})=e^{-i\pi
k\bar{z}-\frac{\pi}{2}k|\tau|-\frac{e|\tau|}{2}(t_+-t_-)}\;^0\chi_2(\bar{z}).
\label{be}\end{equation}
\ren
For $k\geq 1$ we shall use the following ansatz \begin{equation}
\;^0\chi_1(z) \sim
e^{\alpha z^2+\beta z}\vartheta_3(k z+\gamma \mid k\tau),
\label{a124}\end{equation}
where $\vartheta_3(z\mid\tau)$ is the Jacobi's theta function \cite{Mum} and
$\alpha,\beta$ and $\gamma$ are constants to be found from (1.23) and (1.24).
{}From (\ref{a124})
\new{a}
\begin{equation}
\;^0\chi_1(z+1) \sim e^{\beta+2\alpha z+\alpha}\;^0\chi_1(z)
\label{ax}\end{equation}
and
\add
\new{b}
\begin{equation}
\;^0\chi_1(z+\tau) \sim e^{2z\tau\alpha+\alpha\tau^2+\beta\tau-i\pi(2 k
z+2\gamma+
k \tau)}\;^0\chi_1(z).
\label{bx}\end{equation}
\ren
Comparing (1.26) with (1.23), we have
\begin{equation}
\alpha+\beta=\frac{\pi k }{2|\tau|}-\frac{ie}{2}(t_++t_-)+2\pi in,
\label{a127}\end{equation}
\begin{equation}
2\alpha=\frac{\pi k }{|\tau|},
\end{equation}
\begin{equation}
\alpha\tau^2+\beta\tau-i\pi(2\gamma+ k \tau)=\frac{\pi}{2} k
|\tau|-\frac{e|\tau|}{2}(t_+-
t_-)+2\pi im, \end{equation}
where $n$ and $m$ are arbitrary integers. \par Then
\begin{equation}
\alpha=\frac{\pi k }{2|\tau|},
\end{equation}
\begin{equation}
\beta=-\frac{ie}{2}(t_++t_-)+2\pi in
\end{equation}
and
\begin{equation}
\gamma=n\tau-\frac{iet_+}{2\pi}|\tau| - m. \end{equation} Since
$\vartheta_3(z+1\mid\tau)=\vartheta_3(z\mid\tau)$, we get the following
expression
for $ k $ zero modes of the free Dirac equation in the sector with the
topological charge $ k \geq1$ (in accordance with the Index theorem) up to a
normalization factor

\begin{equation}
\;^0\chi_1^{(n)}(z)=e^{\frac{\pi k}{2|\tau|}z^2-\frac{ie}{2}(t_++t_-)z +2\pi
inz}\vartheta_3\left(\left.kz-\frac{iet_+}{2\pi}|\tau|+n\tau\frac{}{}
\right| k\tau\right),
\end{equation}

\(n=0,1,\cdots, k -1\).

Reconstructing $L_1$ and introducing the normalization factor $1/{\cal
N}_{(n)}$ we
obtain from (\ref{aw}) and (\ref{a120}) \begin{equation}
\chi^{(n)}(z,\bar{z})=\frac{1}{{\cal
N}_{(n)}}e^{i\frac{e}{2}x_{\mu}t_{\mu}+\frac{\pi k
}{2|\tau|}(z'^2-z'\bar{z}')+2\pi
inz'}\vartheta_3( k z'+n\tau\mid k \tau) {1 \choose 0},
\label{a134}\end{equation}
where
\begin{equation}
z'=\frac{1}{L_1}(x_1+ix_2)-\frac{ieL_1|\tau|}{2\pi k }t_+. \end{equation}
{}From the
definition of the $\vartheta_3$ function it follows that
$\chi^{(n)}(z,\bar{z})'s$ for
different values of $n$ are orthogonal \begin{equation}
\int_{0}^{L_1}\int_{0}^{L_2}\bar{\chi}^{(n_1)}(z,\bar{z})\chi^{(n_2)}(z,\bar
{z})dx_1dx_
2=0, \end{equation}
if $n_1\neq n_2,$ and in order to get the normalized zero modes one should find
the normalization factor
$1/{\cal N}_{(n)}$.
\par
For the product $\bar{\chi}^{(n)}(z,\bar{z})\chi^{(n)}(z,\bar{z})$ we have from
(\ref{a134})
\begin{equation}
\bar{\chi}^{(n)}(z,\bar{z})\chi^{(n)}(z,\bar{z})=\frac{1}{{\cal
N}^2_{(n)}}\;e^{-\frac{2\pi
Y^2}{|\tau'|}+\frac{2\pi n^2|\tau'|}{ k ^2}}\vartheta_3(X-
iY\mid\tau')\vartheta_3(X+iY\mid\tau'), \end{equation} where
$$ \tau'\equiv k \tau,\;\;\;\;X\equiv \frac{ k x_1}{L_1}+
\frac{eL_1|\tau|}{2\pi}t_2,\;\;\;\;Y\equiv \frac{ k x_2}{L_1}-
\frac{eL_1|\tau|}{2\pi}t_1+n|\tau|. $$

Using relations among the $\vartheta$- functions (see \cite{Whit})

\begin{eqnarray}
\vartheta_3(X-iY\mid\tau')\vartheta_3(X+iY\mid\tau')&=&\frac{1}
{\vartheta^2_4(0\mid\tau')}\{\vartheta_3^2(X\mid\tau')\vartheta_4^2
(iY\mid\tau')
\nonumber \\
&-&\vartheta_2^2(X\mid\tau')\vartheta_1^2(iY\mid\tau')\}. \nonumber
\end{eqnarray}
and
\[ \vartheta_4(iY\mid\tau')=\frac{1}{\sqrt{|\tau'|}}e^{\frac{\pi Y^2}
{|\tau'|}}\vartheta_2\left(\frac{Y}{|\tau'|}\left|-\frac{1}{\tau'}\right)
\right. ,\]
\[\vartheta_1(iY\mid\tau')=\frac{1}{\sqrt{|\tau'|}}e^{\frac{\pi Y^2}
{|\tau'|}}\vartheta_1\left(\frac{Y}{|\tau'|}\left|-\frac{1}{\tau'}\right)
\right..\]
from\cite{Batem} (13.22 (8)) we get
\begin{eqnarray}
\bar{\chi}^{(n)}(z,\bar{z})\chi^{(n)}(z,\bar{z}) & = & \frac{1} {{\cal
N}^2_{(n)}}\frac{e^{\frac{2\pi n^2|\tau'|}{k^2}}}{|\tau'|
\vartheta_4^2(0\mid\tau')}\left\{\vartheta_3^2(X\mid\tau')
\vartheta_2^2\left(\frac{Y}{|\tau'|}\left|
-\frac{1}{\tau'}\right)\right.\right.\nonumber \\
&+&\left.\vartheta^2_2(X\mid\tau')\vartheta_1^2\left(\frac{Y}{|\tau'|} \left|-
\frac{1}{\tau'}\right)\right\}\right. \end{eqnarray} With the help of Landen's
transformation \cite{Batem} (13.23(16)) we can rewrite it
in the following form \par
\begin{eqnarray}
&&\bar{\chi}^{(n)}(z,\bar{z})\chi^{(n)}(z,\bar{z})\nonumber \\ &&=\frac{e^{\frac{2\pi
n^2|\tau'|}{ k ^2}}}{|\tau'|\vartheta_4^2(0\mid\tau') {\cal
N}^2_{(n)}}\left\{\frac{}{}[\vartheta_2(2X\mid 2\tau')\right. \nonumber \\ &&-
\vartheta_3(2X\mid 2\tau')]\vartheta_2\left(\frac{2Y}{|\tau'|} \left|-
\frac{2}{\tau'}\right.\right)(A_2-A_3)B_3 +[\vartheta_2(2X\mid
2\tau')\nonumber \\
&&\left.+\vartheta_3(2X\mid 2\tau')]\vartheta_3\left(\frac{2Y}{|\tau'|}\left|-
\frac{2}{\tau'}\right.\right)(A_2+A_3)B_2\right\}, \end{eqnarray}
 where
$A_i=\vartheta_i(0\mid 2\tau')$,
$B_i=\vartheta_i(0\mid-\frac{2}{\tau'}),$$i=2,3$.
{}From the normalization condition
\begin{equation}
\int_{0}^{L_1}\int_{0}^{L_2}\bar{\chi}^{(n)}(z,\bar{z})
\chi^{(n)}(z,\bar{z}) dx_1dx_2=1
\end{equation}
we get
\begin{eqnarray}
{\cal N}^2_{(n)} & = & \frac{L_1^2 e^{\frac{2\pi n^2|\tau|}{ k }}}{ k |\tau|
\vartheta_4^2(0\mid k \tau)}(A_2+A_3) \nonumber \\ & = & \frac{L_1^2
e^{\frac{2\pi
n^2|\tau|}{ k }}\sqrt{|\tau|}}{\sqrt{2 k }} =\frac{L_1L_2e^{\frac{2\pi
n^2|\tau|}{ k
}}}{\sqrt{2 k |\tau|}}. \label{a141} \end{eqnarray}
Thus the normalized zero modes in the sector with $|k|\geq 1$ are: \\
for $k\geq 1$

\begin{eqnarray}
&&\chi^{(n)}(z,\bar{z})
=\left(\frac{2 k }{|\tau|}\right)^{1/4}\frac{1}{L_1}e^{i\frac{e}
{2}x_{\mu}t_{\mu}+\frac{\pi
k }{2|\tau|}(z'^2-z'\bar{z}')-\frac{\pi n^2|\tau|}{ k }+2\pi inz'}\nonumber\\
&&\times\vartheta_3( k z'+n\tau\mid k\tau) {1 \choose 0}\label{a142}
\end{eqnarray}
\[n=0,1,\cdots, k -1,\;\;\;\;\;\;\;\;\;
z'=\frac{1}{L_1}(x_1+ix_2)-\frac{ieL_1|\tau|}{2\pi k }t_+, \] for $ k \leq -1$

\begin{eqnarray}
&&\phi^{(n)}(z,\bar{z})
=\left(\frac{2|k|}{|\tau|}\right)^{1/4}
\frac{1}{L_1}e^{i\frac{e}{2}x_{\mu}t_{\mu}+\frac{\pi |k|}
{2|\tau|}(\bar{z}''^2-\bar{z}''
z'')-\frac{\pi n^2|\tau|} {|k|}+2\pi n\bar{z}''}\nonumber\\
&&\times\vartheta_3\left(|k|\bar{z}''+n\tau \mid |k|\tau\right) {0 \choose
1} \label{z1}
\end{eqnarray}
\[n=0,1,\cdots, |k|-1;\;\;\;\;\;\;\;\;
\bar{z}''=\frac{1}{L_1}(x_1-ix_2)-\frac{ieL_1|\tau|}{2\pi|k|}t_-. \]

\subsection{Spectrum of the $D_0$ operator} \subsubsection{Trivial sector
$k=0$.}

We have the following eigenvalue and eigenfunction equation for the $\tilde
{D}_0\equiv L_1D_0$ operator \begin{eqnarray} & &\tilde{D}_0^{( k
=0)}\psi(x)\nonumber \\ &=&L_1\left (\begin{array} {cc} 0 & \partial
_1-i\partial_2-
ie(t_1-it_2) \\ \partial_1+i\partial_2-ie(t_1+it_2) & 0
\end{array}
\right ) \psi(x) \nonumber\\
&=&\varepsilon\psi(x).\label{1.44}
\end{eqnarray}
In our case, when $A_{\mu}=t_{\mu}$ and $t_{\mu}$ is a constant, the functions
$\Lambda_{\nu}(x)$ from (\ref{a13}) are constants. So the periodicity
conditions for
the normalized solutions are as follows \begin{equation}
\psi(x+L_{\nu}\hat{\nu})=\psi(x).
\end{equation}
Then one can look for the solution of (\ref{1.44}) using the ansatz
\begin{equation}
\psi(x)=\left (\begin{array}{c}a_1\\a_2\end{array}\right)e^{2\pi
i(\frac{n_1}{L_1}x_1+
\frac{n_2}{L_2}x_2)},
\end{equation}
where $n_{\nu}=0,\pm 1,\pm 2,\cdots$, and $a_1$ and $a_2$ are some complex
numbers which should be found from normalization condition.

We get for the eigenvalues
\begin{equation}
\varepsilon=\pm 2\pi \sqrt{(n_1-\tilde{t}_1)^2+\frac{1}{|\tau|^2}(n_2-
\tilde{t}_2)^2}=\pm i\sqrt{\bar{n}_+\bar{n}_-}, \end{equation} where
$\tilde{t}_{\nu}=\frac{eL_{\nu}}{2\pi}t_{\nu}$, $\bar{n}_{\pm}=2\pi [(n_1\pm
\frac{i}{|\tau|}n_2)-(\tilde{t}_1\pm \frac{i}{|\tau|}\tilde{t}_2)]$.

Using the normalization condition
\begin{equation}
\int_{0}^{L_1}dx_1\int_{0}^{L_2}dx_2 \bar{\psi}(x)\psi(x)=1, \end{equation}
we get,
that $a_1$ and $a_2$ can be any complex numbers subject the
condition:$|a_1|^2+|a_2|^2=\frac{1}{L_1L_2}$.

\subsubsection{Nontrivial sector $k\neq 0$.}

Using the form (\ref{a118}) for the vector potential in the case $ k \neq0$ and
complex variables, we get (for dimensionless operator) \begin{eqnarray}
\tilde{D}_0^{( k \neq0)} & \equiv & L_1D_0^{( k \neq0)}=2\left(
\begin{array}{cc}
0 & \partial_z-\frac{eBL_1^2}{4}\bar{z}-\frac{ieL_1}{2}t_- \\
\partial_{\bar{z}}+\frac{eBL_1^2}{4}z-\frac{ieL_1}{2}t_+ & 0 \end{array}
\right)
\nonumber \\
& = & \sqrt{2e|B| L_1^2}\left (\begin{array}{cc} 0 & d_+ \\ d_- & 0
\end{array} \right ),
\end{eqnarray}
where $B\equiv \frac{2\pi k }{eL_1L_2}$ , $d_+\equiv \sqrt{\frac{2}{e|B|
L_1^2}}(\partial_z-\frac{eBL_1^2}{4}\bar{z}-\frac{ieL_1}{2}t_-)$ \newline
and $d_-
\equiv \sqrt{\frac{2}{e|B| L_1^2}} (\partial_{\bar{z}}+\frac{eBL_1^2}{4}z-
\frac{ieL_1}{2}t_+)$. Note that $(d_+)^{\dagger}=-d_-$ and \begin{equation}
[d_+,d_-]=\frac{k}{|k|}\label{a1.50}
\end{equation}
Then
\begin{equation}
\tilde{D}_0^{\dagger}=-\sqrt{2e|B| L_1^2}\left ( \begin{array}{cc} 0 & d_+ \\
d_- & 0
\end{array}
\right )
\end{equation}
and
\begin{equation}
\tilde{D}_0^{\dagger}\tilde{D}_0=-2e|B| L_1^2 \left ( \begin{array}{cc}
d_+d_- & 0 \\ 0
& d_-d_+ \end{array} \right ) . \end{equation} \par
Let us consider the cases of positive and negative $ k $ separately.

{\bf A.} $k>0$.

First let us define the vacuum states in the Fock space with creation and
annihilation
operators $d_+$ and $d_-$ functions, which are solutions of the equation
\begin{equation}
d_-\chi_1(z,\bar{z})=0
\label{a153}\end{equation}
We already know that there should be exactly $ k $ such functions. Equation
(\ref{a153}) is the equation for the first component of the spinor
$\chi(z,\bar{z})$ in
(\ref{a120}) with the normalized solutions given in (\ref{a142}).

Then the eigenfunctions of the $\tilde{D}_0^{\dagger}\tilde{D}_0$ operator
are the
functions \begin{equation}
\left (\begin{array}{c}
(d_+)^m\chi_1^{(n)}(z,\bar{z})\\
0
\end{array} \right )\;\;\;\makebox{and}\;\;\;\left( \begin{array}{c} 0 \\
(d_+)^{m-
1}\chi_1^{(n)}(z,\bar{z}) \end{array} \right ), \end{equation} which
correspond to
eigenvalue $\varepsilon^2_m=\frac{4\pi k}{|\tau|}m$, where $m=0,1,2,\cdots$,
$n=0,1,2,\cdots,k-1$.

Let us check this for $m>0$
\begin{eqnarray}
& &\tilde{D}_0^{\dagger}\tilde{D}_0 \left(\begin{array}{c}
(d_+)^m\chi_1^{(n)}(z,\bar{z})\\
0
\end{array}\right ) \nonumber \\
&&= -\frac{4\pi k }{|\tau|}\left (\begin{array}{cc} d_+d_- & 0 \\ 0 &
d_-d_+ \end{array}
\right )\left(\begin{array}{c}
(d_+)^m\chi_1^{(n)}(z,\bar{z}) \\
0 \end{array} \right ) \\
&&= -\frac{4\pi k }{|\tau|}\left (\begin{array}{c}
d_+d_-(d_+)^m\chi_1^{(n)} (z,\bar{z})\\
0 \end{array} \right )
\nonumber
\end{eqnarray}
{}From (\ref{a1.50}) we have $d_-(d_+)^m=(d_+)^md_--m(d_+)^{m-1}$. Then
\new{a}
\begin{equation}
\tilde{D}_0^{\dagger}\tilde{D}_0\left(\begin{array}{c}
(d_+)^m\chi_1^{(n)}(z,\bar{z})\\
0 \end{array} \right )=\frac{4\pi k m}{|\tau|} \left ( \begin{array}{c}
(d_+)^m\chi_1^{(n)}(z,\bar{z}) \\ 0 \end{array} \right ). \end{equation}
Similarly
\add
\new{b}
\begin{eqnarray}
&&\tilde{D}_0^{\dagger}\tilde{D}_0\left (\begin{array}{c} 0 \\ (d_+)^{m-
1}\chi_1^{(n)}(z,\bar{z}) \end{array}\right ) \nonumber \\ &&= -\frac{4\pi
k }{|\tau|}\left (
\begin{array}{c} 0 \\ (d_-d_+)(d_+)^{m-1}\chi_1^{(n)}(z,\bar{z})
\end{array} \right ) \\
&&= \frac{4\pi k m}{|\tau|}\left(\begin{array}{c} 0 \\
(d_+)^{m-1}\chi_1^{(n)}(z,\bar{z})
\end{array} \right ). \nonumber
\end{eqnarray}
\ren
Since the index $i$ takes $ k $ values, the spectrum of the
$\tilde{D}_0^{\dagger}\tilde{D}_0$ operator has a $2k$ fold degeneracy for
$m>0$,
and a $k$ fold degeneracy for $m=0$.

One should remember that $\tilde{D}_0^{\dagger}=-\tilde{D}_0$, so for the
eigenvalues of the $\tilde{D}_0$ operator we have: $\varepsilon=0\;,\pm
i\sqrt{\frac{4\pi k m}{|\tau|}}$ and each value has a $ k $-fold degeneracy.

The eigenfunctions of this operator have the following form $(k>0)$ \new{a}
\begin{equation}
\psi^{(n)}_{+,m}(x)=\frac{1}{{\cal N}^{(n)}_{m,+}}\left ( \begin{array}{c}
(d_+)^m\chi_1^{(n)}(z,\bar{z}) \\
i\sqrt{m}(d_+)^{m-1}\chi_1^{(n)}(z,\bar{z}) \end{array} \right )
\end{equation} for
$\varepsilon=+i\sqrt{\frac{4\pi k m}{|\tau|}}$, and \add \new{b}
\begin{equation}
\psi^{(n)}_{-,m}(x)=\frac{1}{{\cal N}^{(n)}_{m,-}}{
(d_+)^m\chi_1^{(n)}(z,\bar{z})
\choose
-i\sqrt{m}(d_+)^{m-1}\chi_1^{(n)}(z,\bar{z}) } \end{equation} \ren
for $\varepsilon=-i\sqrt{\frac{4\pi k m}{|\tau|}}\;,1/{\cal
N}^{(n)}_{m,\pm}$ is a
normalization factor.

{\bf B.} $k<0$.

\ren
Now the vacuum state is defined as follows \begin{equation}
d_+\phi_2(z,\bar{z})=0
\end{equation}
There are $|k|$ vacuum states in accordance with the index theorem. They are
given in (\ref{z1}).
\par
The eigenvalues of the $\tilde{D}_0^{\dagger}\tilde{D}_0$ operator
$\varepsilon^2_m=-\frac{4\pi k m}{|\tau|}$ are $2|k|$ fold degenerate for
$m>0$, and
$|k|$ fold degenerate for $n=0$, and its eigenfunctions are as follows
\begin{equation}
\left ( \begin{array}{c}
0 \\
(d_-)^m\phi_2^{(n)}
\end{array} \right ),\;\;\;\;\;\;\left ( \begin{array}{c}
(d_-)^m\phi_2^{(n)} \\ 0 \end{array}
\right )\;\;\;\;\makebox{for}\;\;m>0. \end{equation} In this case, for the
eigenvalues of
the $\tilde{D}_0$ operator we have: $\varepsilon=0,\pm i\sqrt{\frac{4\pi|k|
m}{|\tau|}}$, $m=1,2,\cdots$, with a $|k|$ fold
degeneracy and corresponding eigenfunctions $(n=0,\cdots,|k|-1)$ \new{a}
\begin{equation}
\tilde{\psi}^{(n)}_{+,m}(x)=\frac{1}{\tilde{{\cal N}}^{(n)}_{m,+}}\left
(\begin{array}{c}
i\sqrt{m}(d_-)^{m-1}\phi^{(n)}_2(z,\bar{z})\\
(d_-)^m\phi_2^{(n)}(z,\bar{z}) \end{array}
\right ) \end{equation}
for $\varepsilon=i\sqrt{\frac{4\pi|k|}{|\tau| m}}$, and \add \new{b}
\begin{equation}
\tilde{\psi}^{(n)}_{-,m}(x)=\frac{1}{\tilde{{\cal N}}^{(n)}_{m,-}}\left
(\begin{array}{c}
i\sqrt{m}(d_-)^{m-1}\phi^{(n)}_2(z,\bar{z})\\
-(d_-)^m\phi_2^{(n)}(z,\bar{z}) \end{array}
\right ) \end{equation}
for $\varepsilon=-i\sqrt{\frac{4\pi|k|}{|\tau| m}}$.

\renewcommand{\ren}{\renewcommand{\theequation}{2.\arabic{equation}}}
\renewcommand{\add}{\addtocounter{equation}{-1}}
\renewcommand{\new}[1]{\renewcommand{\theequation}{2.\arabic{equation}#1}}
\newcommand{\set}{\setcounter{equation}{0}}

\set
\section{Quantum theory}

The calculation of the quantum mechanical vacuum expectation values (VEV) of
observables $\Omega(A,\bar{\psi},\psi)$ is performed with the help of the path
integral formula

\ren
\begin{equation}
\langle \Omega (A,\bar {\psi},\psi )\rangle=\frac{1}{Z} \int{\cal D}A{\cal
D} \bar
{\psi}{\cal D}\psi e^{-S[A,\bar {\psi},\psi ]}\Omega (A,\bar {\psi} ,\psi).
\end{equation}
Since the action (1.1) is quadratic in fermion fields the fermionic
integration can easily be
done.
Recollecting that there are different topological sectors of gauge field
configurations
we have
\begin{eqnarray}
\langle\Omega (A,\bar {\psi},\psi )\rangle&=&\frac{1}{Z}\sum_{k=-\infty
}^{\infty}
\int_{{\cal A}_k}{\cal D}A{\cal D}\bar{\psi}{\cal D}\psi
e^{-S[A_{\mu}]-\int_{{\cal T}}
d^2x\bar{\psi}D\psi}\nonumber \\
&\times& \Omega(A,\bar{\psi},\psi),
\end{eqnarray}
where the partition function

\begin{equation}
Z=\sum_{k=-\infty }^{\infty} \int_{{\cal A}_k}{\cal D}A{\cal D}\bar{\psi}
{\cal D}\psi e^{-
S[A_{\mu}]-\int_{{\cal T}} d^2x\bar{\psi}D\psi}. \end{equation}

The result of the fermionic integration depends crucially on the number
$|k|$ of zero
modes of the operator $D[A]$ dependent of the gauge field $A_{\mu}$ from the
topological sector ${\cal A}_k$ \cite{Camil} and we obtain \begin{equation}
\langle\Omega (A)\rangle=Z^{-1} \int_{{\cal A}_0}{\cal D}A\det\tilde {D} e^{-
S[A_{\mu} ]}\Omega (A)
\end{equation}
\begin{eqnarray}
&&\langle\psi_{\alpha_1}(x_1)\bar{\psi}_{\beta_1}(y_1)\cdots\psi_{\alpha_N}
(x_N)\bar{\psi}_{\beta_N}(y_N)\rangle \nonumber \\ &&=Z^{-1}\sum_{k=0\pm1,\cdots
,\pm N}L_1^{|k|}\int_{{\cal A}_k}{\cal D}Ae^{-S[A]} {\det}'[\tilde D]
\label{q11} \\
&&\times\sum_{P}(-1)^p\chi_{\alpha_{1}}^{(1)}(x_{i_1})\cdots
\chi_{\alpha_{|k|}}^{(|k|)}
(x_{i_{|k|}})\bar{\chi}_{\beta_1}^{(1)}(y_{j_1})\cdots
\bar{\chi}_{\beta_{|k|}}^{(|k|)}(y_{j_{|k|}}) \nonumber \\ &&\times
S_{{\alpha}_{|k|+1}{\beta}_{|k|+1}}^{(k)} (x_{i_{|k|+1}},y_{j_{|k|+1}};A)\cdots
S_{{\alpha}_N {\beta}_N}^{(k)}(x_{i_N},y_{j_N};A),\nonumber \end{eqnarray}
\begin{equation}
Z=\int _{{\cal A}_0}{\cal D}Ae^{-S[A]}\det[\tilde{D}], \end{equation}
$(\chi^{(n)}(x),\bar{\chi}^{(n)}(y)\;\;\;\;n=1,\cdots ,|k|)$ is an
orthonormal set of the zero
mode wave functions and $S^{(k)}(x,y;A)$ is a Green's function of the $D[A]$
operator with a gauge field $A_{\mu}$ from the topological sector ${\cal A}_k$,
${\det}'[\tilde{D}]$ is a product of nonzero eigenvalues of the dimensionless
$\tilde{D}= L_1D$ operator. The sum is taken over all possible permutations
$P=(i_1,\cdots ,
i_N)$ of $(1,\cdots,N),(-1)^p$ is a parity of the permutation $P$.\footnote
{Note that
$[\chi]=[\bar{\chi}]=[S^{(k)}]=[l]^{-1}$.}

All these expressions are formal and need regularization. We will use the
Pauli-Villars
regularization. It means that we will add to the action a sum \begin{equation}
\sum_{i=1}^{r}\int d^2x \bar{\psi}_i(D-M_i)\psi_i , \end{equation} where
the regulator
masses $M_i>0, \; i=1,\cdots,r$ satisfy the regulator conditions
\newline
$\sum _{i=1}^r e_i(M_iL_1)^{2p}=0,\;\;\;p=1,\cdots, r-1, \sum_{i=1}^re_i=-1$ and
$e_i=+1(-1)$, if $\psi_i$ is a fermion (boson) field, $r$ is a number of
regular fields,
sufficient to regularize all singularities which appeared in the theory
\cite{LILEU}.

In what follows we will use only dimensionless operators $\tilde{D},
\tilde{D}\tilde{D}^{\dagger}$ etc. and will not write tilde. Then
\begin{equation}
{\det}'D \to{\det}'D \prod_{i=1}^r
\det(D-M_iL_1)^{e_i}\equiv\exp\frac{1}{2}\Gamma_{reg}^{(k)}[A]. \label{q1}
\end{equation}
It is more convenient to work with the operator $DD^{\dagger}$ which has a
nonnegative spectrum. Then (see Appendix {\bf A}) \begin{equation}
\Gamma _{reg}^{( k )}[A]=2\pi i|k|+ {\rm Tr}'\ln DD^{\dagger}+{\rm Tr}
\sum_{i=1}^{r}
e_i \ln(DD^{\dagger} +M_i^2 L_1^2) \label{q2} \end{equation}
The prime on the trace means that the zero modes are excluded. The
regularization
allows the following mathematical operations leading to an evaluation of the
effective action $\Gamma _{reg}^{( k )}[A]$ .

a) First, we will find the variation of $\Gamma _{reg}^{(k)}[A]$ under a
variation of
$b(x)$ in the expression of the potential $A_{\mu}$ (see (1.9)), then by
solving the
variational equation we will obtain $\Gamma _{reg}^{(k)}[A]$ in the
following form
\begin{equation}
\Gamma _{reg}^{(k)}[A]= F[b]+\alpha( k ,M_iL_1),\label{q7} \end{equation} where
$\alpha(k,M_iL_1)$ does not depend on $b(x)$.

b) Second, we will calculate $\alpha(k,M_iL_1)$.

The regularized action (\ref{q2}) may be rewritten as follows \begin{eqnarray}
\Gamma_{reg}^{(k)}[A]&=&2\pi i|k|+|k|\sum_{i=1}^{r} e_i\ln(M_i^2 L_1^2)
\nonumber
\\
&-&\int_0^{\infty}\frac{dt}{t}({\rm Tr} e^{-tDD^{\dagger}}- |k|)\left(1+\sum
_{i=1}^{r}e_ie^{-tM_i^2L_1^2}\right). \end{eqnarray} Therefore under the
variation
of $b(x)$ we get \begin{equation} \delta \Gamma_{reg}^{( k )}[A]=-
\int_0^{\infty}\delta({\rm Tr}e^{-tDD^{\dagger}})
\left(1+\sum_{i=1}^{r}e_ie^{-tM_i^2L_1^2}\right)\frac{dt}{t}. \label{q3}
\end{equation}
In Appendix {\bf B} it is shown that \begin{equation} \delta({\rm Tr} e^{-
tDD^{\dagger}})=4et\frac{d}{dt}{\rm Tr}(\gamma_5\delta be^{-
tDD^{\dagger}}).\label{q4}
\end{equation}
Substituting this formula into (\ref{q3}), we have \begin{eqnarray} \delta
\Gamma
_{reg}^{( k )}[A]&=&\left. -4e{\rm Tr}(\gamma_5\delta be^{- tDD^{\dagger}})
\left(1+\sum_{i=1}^{r}e_ie^{-
tM_i^2L_1^2}\right)\right|_{0}^{\infty}\nonumber \\ &-
&4e\sum _{i=1}^{r}e_iM_i^2L_1^2\int _0^{\infty}dt {\rm Tr}(\gamma_5\delta be^{-
tDD^{\dagger}})e^{-tM_i^2L_1^2}. \label{q5} \end{eqnarray} The first term in
(\ref{q5}) contributes only on the upper limit, if the $DD^{\dagger}$
operator has zero modes. So when $t\to\infty$ \begin{equation} {\rm
Tr}(\gamma_5
\delta be^{-tDD^{\dagger}})\to {\rm Tr}(\gamma_5\delta b{\cal P}_0),
\end{equation}
where ${\cal P}_0$ is the projection operator on the subspace spanned by
the zero
modes.
To calculate the second term we will use the heat kernel expansion of $e^{-tDD^
{\dagger}}$ for small $t$ \cite{Gil}.
\begin{equation}
\langle x\mid e^{-tDD^{\dagger}}\mid x\rangle=\frac{1}{4\pi t}[1+e\gamma_5
F_{12}
(x)t+\cdots ]
\end{equation}
and for $M_i\to\infty$
\begin{equation}
\int_0^{\infty} dt {\rm Tr}(\gamma_5 \delta be^{-tDD^{\dagger}}) e^{-
tM_i^2L_1^2}=\frac{1}{2\pi M_i^2L_1^2}\int_{\cal T}d^2 x\delta beF_{12}(x).
\end{equation}
Finally we obtain
\begin{eqnarray}
\delta\Gamma_{reg}^{( k )}[A]&=&-4e{\rm Tr}(\gamma_5 \delta b {\cal P}_0)- \frac
{2e^2}{\pi}\int_{\cal T}d^2 x\delta b(x)F_{12}(x)\nonumber \\ &=&\delta
2(\ln\det{\cal
N}_A^{( k )})+\delta\left(\frac {e^2}{\pi} \int_{\cal T}d^2 x b(x)
\Box b(x)\right),\label{q6} \end{eqnarray} where , $ \Box =\partial_1^2
+\partial_2^2
$ and ${\cal N}_A^{( k )}$ is a $k\times k$
matrix of the scalar products of the zero modes $${\cal
N}_A^{(k)}=||(\hat{\chi}^{(n)},\hat{\chi}^{(n')})||,\;\;n,n'=0,\cdots,k-1.$$
The variational
equation (\ref{q6}) has the solution \begin{equation} \Gamma_{reg}^{( k
)}[A]=2\ln \;
\det{\cal N}_A^{(k)}+\frac{e^2}{\pi} \int_{\tau} d^2xb(x)
\Box b(x)+\alpha (k,M_iL_1), \end{equation} where $\alpha(k,M_iL_1)$ is the
integration constant.\\ Now let us find $\alpha(k,M_iL_1)$. From (\ref{q2}) and
(\ref{q7}) it follows that \begin{eqnarray}
\alpha(k,M_iL_1)&=&\Gamma_{reg}^{(k)}[A]\mid_{b=0}=2\pi i|
k|+|k|\sum_{i=1}^{r}e_i\ln M_i^2L_1^2 \nonumber \\ &+& {\rm Tr}^{\prime}\left
\{\sum_{i=1}^{r}e_i\ln(D_0D_0^{\dagger}+M_i^2L_1^2)+ \ln(D_0D_0^{\dagger})
\right \}.\label{q8} \end{eqnarray}

To calculate the trace in the second line of (\ref{q8}) we use the
L\"uscher's formula
(see Appendix {\bf C}) and
\begin{eqnarray}
& &{\rm Tr}'\left\{\sum _{i=1}^{r}e_i \ln(D_0D_0^{\dagger}+M_i^2L_1^2)
+\ln(D_0D_0^{\dagger}) \right\} \label{q10} \\ & &=\sum_{i=1}^{r} e_i\left(-
r_1(D_0D_0^{\dagger})M_i^2 L_1^2 \ln M_i^2L_1^2 +\zeta(0\mid
D_0D_0^\dagger)\ln M_i^2L_1^2\right)-\zeta'(0\mid D_0D_0^\dagger), \nonumber
\end{eqnarray}
where terms vanishing for $M_i\rightarrow \infty $ have been neglected, $\zeta
(s)\equiv \zeta (s\mid D_0D_0^\dagger)$ is a $\zeta$-function of the
$D_0D_0^\dagger$ operator
\begin{equation}
\zeta(s\mid D_0D_0^{\dagger})=\sum_{m}(\lambda_m)^{-s}d_m \label{q9}
\end{equation}
is defined for sufficiently large real part Re$s$ and a sum is taken over
all nonzero
eigenvalues $\lambda_m$ of $D_0D_0^{\dagger}$ operator with multiplicity
$d_m$. This function has a pole at $s=1$ with the residue
$r_1(D_0D_0^{\dagger})$
\begin{equation}
\zeta (s \mid D_0D_0^{\dagger})=\frac {r_1(D_0D_0^{\dagger})}{s-1}+\cdots.
\label{q12}
\end{equation}
Thus we see, that in order to find $\Gamma _{reg}^{(k)}[A]$, we should know
$r_1(D_0D_0^{\dagger}), \;\; \zeta (0 \mid D_0D_0^{\dagger})$ and $\zeta'
(0 \mid
D_0D_0^{\dagger})$. But the spectrum of the $D_0D_0^{\dagger}$ operator or the
eigenvalues $\lambda_m$ and their multiplicity is known and after some
calculations connected with summation of infinite series (see Appendix {\bf
D}) we
get the final result \begin{eqnarray}
& & \Gamma_{reg}^{(k)}[A]=\frac{e^2}{\pi}\int_{\cal T}d^2xb(x)\Box b(x)
+2\ln\det{\cal
N}_A^{(k)}-|k|\ln\frac{2|k|}{|\tau|} \nonumber \\ & &+ \delta_{0 k
}\left\{2\ln|\vartheta_1(t\mid\tau)\eta^{-1}(\tau)|^2-
\frac{4\pi}{|\tau|}\bar{t}_1^2\right
\}\label{2.24} \\ &+& \sum_{j=1}^{r}e_j\left(-\frac{|\tau|M_j^2L_1^2}{2\pi} \ln
(M_j^2L_1^2)+2\pi i|k|\right).\nonumber \end{eqnarray}

We want to discuss shortly the implications of this result for the path
integral formula
after fermion integration, Eq.(\ref{q11}). The normalization factor $Z$
follows from
the condition $\langle 1\rangle =1$ and therefore is determined by the
integration
over the trivial sector $ k =0$ only. The $k$ independent parts of
$\Gamma_{reg}^{(k)}[A]$ can be factorized out of the sum over
the gauge sectors. These are the regulator mass dependent term and the term with
$b(x)$. Therefore the regulator mass term in $\Gamma_{reg}^{( k )}[A]$ cancels
against that of $Z$ and the normalized result is independent of these masses as
required for a consistent renormalization scheme. Also the integration over
the pure
gauge component ${\cal D}a$ factorizes from all other integrations. Thus we
let this
contribution cancel with the corresponding of $Z$ without going into
further details
of gauge fixing.

After fermion integration the gauge field dependent part of the action in
the trivial
sector is

\begin{eqnarray}
S'[b(x)]&=&\frac{1}{2}\int_{\cal T}d^2 x \left (F_{12}(x)F_{12}(x)-\frac
{e^2}{\pi}
b(x)\Box b(x) \right ) \nonumber \\ &=&\frac {1}{2}\int_{\cal T}d^2 x b(x)\Box (\Box
-m^2)b(x), \end{eqnarray}
where $m^2=\frac{e^2}{\pi}$. As we see this action is bilinear in $b(x)$
and hence
describes free particles. The generating functional of its correlation
functions can be
calculated by Gaussian integration: \begin{equation} \frac {1}{Z_0}\int
{\cal D}be^{-
S'[b]}e^{\int _{\cal T}d^2 x J(x)b(x)}=e^{\frac {1} {2}\int_{\cal T}d^2 x
\int_{\cal T}d^2 x'
J(x)G(x-x')J(x')} \end{equation} and
\begin{equation}
Z_0=\int{\cal D}be^{-S'[b]} .
\end{equation}
The propagator $G(x-x')$ related to $S'[b]$ obeys the following equation
\begin{equation}
\Box(\Box-m^2)G(x-x')=\delta^{(\prime)}(x-x')\equiv\delta(x-x')-\frac{1}{L_1
L_2}
\end{equation}
and it plays an imporatant role in future calculations. We may express $G(x)$ by
the eigenfuctions and eigenvalues of the Laplacian $\Box$ on ${\cal T}$:
\begin{eqnarray}
G(x)&=&\frac{1}{m^2L_1L_2}{\sum_{n_i}}'\left\{\frac{e^{\frac{2\pi i}
{L_1}(n_1x_1+|\tau|^{-1}n_2x_2)}}{(\frac{2\pi}{L_1})^2(n_1^2+|\tau |^{-
2}n_2^2)}\right. \nonumber \\
&-&\left.\frac{e^{\frac{2\pi i}{L_1}(n_1x_1+|\tau|^{-1}n_2x_2)}} {m^2
+(\frac{2\pi
}{L_1})^2(n_1^2+|\tau|^{-2}n_2^2)}\right\} \equiv \frac {1}{m^2}\{G_0(x)-
G_m(x)\}.\label{2.29} \end{eqnarray}
The summation $\sum'_{n_i}$ excludes $n_1=n_2=0$. Therefore $G_0(x)$ is the
Green's function of the Laplacian on ${\cal T}$, which as an integral operator
transforms a constant function into zero $G_0\ast(const)=0$. It has a
representation
(see Appendix E ) \begin{eqnarray}
G_0(x)&=&\frac{1}{L_1L_2}{\sum_{n_i}}' \frac {e^{\frac {2\pi
i}{L_1}(n_1x_1+ |\tau|^{-
1}n_2x_2)}}{(\frac {2\pi}{L_1})^2(n_1^2+|\tau|^{-2}n_2^2)} \nonumber \cr
&=&-\frac
{1}{2\pi}\ln\left|\frac{\vartheta_1(z\mid\tau)} {\vartheta'_1
(0\mid\tau)}\right|
+\frac{x_2^2}{2L_1L_2}+\frac{|\tau|}{12}-\frac {1}{2\pi}\ln 2\pi
q_0^2(\tau) \nonumber
\\
&=&-\frac{1}{2\pi}\ln\left (2\pi\eta^2(\tau)e^{-\frac{\pi x^2_2} {L_1L_2}}
\left|\frac{\vartheta_1(z\mid\tau)}{\vartheta'_1(0\mid\tau)}\right|\right).\
\label{2.30}
\end{eqnarray}
Note that $\vartheta_1'(0|\tau)=2\pi \eta^3(\tau)$. With the help of the
formula
\begin{eqnarray}
\sum_{n=-\infty}^{\infty}\frac{e^{2\pi inx}}{a^2+n^2}=\frac{\pi \cosh(\pi
a[1-2|x|])} {a
\sinh(\pi a)}\label{2.31}
\end{eqnarray}
we can do one summation in $G_m(x)$

\begin{eqnarray}
G_m(x)+\frac{1}{L_1L_2m^2}=\bar{G}_m(x)
&=&\sum _{n}\frac {e^{\frac {2\pi i}{L_1}nx_1}\cosh [|\tau|
\xi(n)(1/2-x_2/L_2)]}{2\xi
(n)\sinh [|\tau|\xi (n)/2]} \nonumber \\
&{\mathop{=}\limits_{L_2\to\infty}}&\sum_{n}\frac{e^{\frac {2\pi
i}{L_1}nx_1 -\xi
(n)\frac {x_2}{L_1}}}{2\xi (n)} \label{2.32} \\
&{\mathop{=}\limits_{L_1\to\infty}}&\frac{1}{2\pi}K_0(m|x|)=\frac
{1}{4\pi^2}\int\int\frac{d^2pe^{ipx}}{p^2+m^2}, \nonumber \end{eqnarray}
where $\xi
(n)=\sqrt {4\pi ^2n^2+L_1^2m^2}$ and $\frac {1}{2\pi} K_0(m|x|)$ is the
Green's function of the free particle with a mass $m$ in the infinite two
dimensional
Euclidean space-time (to find the limit $L_1 \to\infty $ we used Euler's
summation
formula) which has the following asymptotics \cite{Abram} \begin{equation}
K_0(m|x|)\rightarrow -\left [\ln\frac{m|x|}{2}+\gamma\right ],
\;\;\;\;\;\;|x|\to 0 \label {2.33}
\end{equation}
\begin{equation}
K_0(m|x|)\to\sqrt{\frac{\pi}{2m|x|}}e^{-m|x|}\left[1+O\left(\frac{1}
{m|x|}\right)\right],\;\;\;\;\;\;|x|\to\infty \label {2.34} \end{equation}

\renewcommand{\theequation}{3.\arabic{equation}} \def\be{\begin{equation}}
\def\ee{\end{equation}}
\def\beq{\begin{eqnarray}}
\def\eeq{\end{eqnarray}}

\set
\section{Applications}

\subsection{Average of fermion bilinears}

Now we calculate the vacuum expectation value of the gauge invariant fermion
bilinear, using the general formula (\ref{q11}) (we do gauge invariant
point-splitting)

\beq
&&\langle M(x)\rangle \equiv\langle \bar\psi(x+\zeta)\Gamma
e^{ie\int\limits_{x-
\zeta}^{x+\zeta}A_\alpha(y)dy_\alpha} \psi(x-\zeta)\rangle\nonumber \\
&&=\frac{1}{Z}\sum_k\int_{{\cal A}_k}{\cal D}\bar\psi{\cal D}\psi{\cal D}A
e^{-S[A]-
S_F[\bar\psi,\psi,A]}\bar\psi(x+\zeta)\tilde\Gamma\psi(x-\zeta)\nonumber\\
&&=\frac{1}{Z}\sum_k\int_{{\cal A}_k}{\cal D}Ae^{-S[A]}I_{(k)}
[x,\zeta;\Gamma],\label{3.1}
\eeq
where $\tilde\Gamma=\Gamma e^{ie\int\limits_{x-\zeta}^{x+\zeta}A_\alpha(y)
dy_\alpha}$,
\begin{eqnarray}
\sum_k\int_{{\cal A}_k}{\cal D}Ae^{-S[A]}...= \sum_k e^{-\frac{2\pi
k^2}{m^2L_1L_2}}\int\limits_{0}^{T_1}dt_1\int\limits_{0}^{T_2} dt_2\int
{\cal D}be^{-
\frac{1}{2}\int b(x)\Box^2b(x)dx}...,\label{3.2} \end{eqnarray} and
$T_{\mu}=2\pi/eL_{\mu}$.

For fermion bilinears only gauge sectors $C{\cal H}^{(k)}$, with $k=0,\pm 1$
contribute.

For the trivial sector ($k=0$):
\be
I_{(0)}[x,\zeta,\Gamma]=-T_\mu[a,b,x,\zeta;\Gamma]S_t^{(\mu)}(-2\zeta)
e^{\frac{1}{2}\Gamma_{reg}^{(0)}[A]}
\ee
and
\beq
&&T_\mu[a,b,x,\zeta;\Gamma]\nonumber\\
&&\equiv e^{-ie\Delta a(x)}[{\rm Tr}(\gamma_\mu\Gamma)\cosh e\Delta b(x)+ {\rm
Tr}(\gamma_5\gamma_\mu\Gamma)\sinh e\Delta b(x)].\label{3.4} \eeq

For the sector $|k|=1$
\be
I_{(1)}[x,\zeta;\Gamma]=-L_1\bar{\hat\chi}(x+\zeta)\Gamma\hat\chi(x-\zeta)
e^{\frac{1}{2}\Gamma_{reg}^{(+1)}[A]},
\ee

\be
I_{(-1)}[x,\zeta;\Gamma]=-L_1\bar{\hat\phi}(x+\zeta)\Gamma\hat\phi(x-\zeta)
e^{\frac{1}{2}\Gamma_{reg}^{(-1)}[A]},
\ee
where $\hat\chi(\hat\phi)$ is a zero mode of positive (negative) chirality
of the
operator $\gamma^\mu(\partial_\mu-ie(t_\mu+\epsilon_{\mu
\nu}\partial_{\nu}b+C^{(\pm 1)}_\mu))$ (see (section 1)).

We see that for the vector ($\Gamma=\gamma_\mu$) or pseudovector
($\Gamma=i\gamma_5\gamma_\mu$) currents there is the contribution only from
the trivial sector, and for scalar ($\Gamma=\bf 1$) or pseudoscalar
($\Gamma=\gamma_5$) there is the contribution only from the topological sectors
$|k|=1$.

So we get, using (\ref{3.1}) and (\ref{f10}) \beq && \langle
\bar\psi(x+\zeta)\gamma_\alpha e^{ie\int_{x-\zeta}^{x+\zeta}A_\beta(y)
dy_\beta}\psi(x-\zeta)\rangle \\
&&{\mathop{\sim}\limits_{\zeta\to
0}}\;\frac{2}{\pi}e^{I_2(\zeta)}[\delta_{\mu\alpha} \cos
I_1(\zeta)+\epsilon_{\mu\alpha}\sin I_1(\zeta)]\nonumber\\
&&\times[\frac{\zeta_\mu}{2|\zeta|^2}-\langle K_\mu(t)\rangle ^{(0)}_t],
\eeq where
\be
I_1(\zeta)\equiv e^2\epsilon_{\alpha\beta}\int_{-\zeta}^\zeta [\partial_\beta
G(y+\zeta)-\partial_\beta G(y-\zeta)]dy_\alpha, \ee

\be
I_2(\zeta)\equiv\frac{e^2}{2}\epsilon_{\alpha\beta}\epsilon_{\gamma\nu} \int_{-
\zeta}^\zeta\int_{-\zeta}^\zeta\partial_\beta\partial_\nu G(x-y)dx_\alpha
dy_\gamma,
\ee
$G(x)$ is the propagator of the $b$-field (\ref{2.29}) and $\langle
\ldots\rangle
_t^{(0)}$ is the averaging over the toron configurations

\be
\langle\ldots\rangle^{(0)}_t=\frac{1}{Z_{toron}}\int_0^{T_1}dt_1\int_0^{T_2}
dt_2e^{\frac{1}{2}\Gamma^{(0)}(t)}\ldots,\label{3.11} \ee and
\beq
&&Z_{toron}=\int_0^{T_1}dt_1\int_0^{T_2}dt_2e^{\frac{1}{2}\Gamma^{(0)}(t)}\
nonumber\\
&&=\int_0^{T_1}dt_1\int_0^{T_2}dt_2|\vartheta_1(i\bar{t}_+|\tau)|^2 e^{-
\frac{2\pi}{|\tau|}\bar{t}^2_1}=
\frac{(2\pi)^2}{e^2\sqrt{2|\tau|}L_1L_2}.\label{3.12} \eeq Using the relation
$K_\mu(t)=\frac{i\pi}{2eL_1L_2}\frac{\partial} {\partial
t_\mu}\Gamma^{(0)}(t)$ (see
(\ref{f11})-(\ref{f12})), one can easily prove that\newline
$\langle K_\mu(t)\rangle _t^{(0)}=0$.

We have the singular behavior for the current \be \langle
\bar\psi(x+\zeta)\gamma_\alpha
e^{ie\int\limits_{x-\zeta}^{x+\zeta}A_\beta(y) dy_\beta}\psi(x-
\zeta)\rangle{\mathop{\sim}\limits_{\zeta\to 0}}
\frac{\zeta_\alpha}{\pi|\zeta|^2}+o(|\zeta|). \ee For the scalar
($\Gamma={\bf 1}$) in
(\ref{3.1}) (in the limit $\zeta\to 0$) , using
(\ref{3.2}),(\ref{2.24}) and (\ref{3.11}) we get \beq \langle
\bar\psi(x)\psi(x)\rangle
&=& 2\frac{\eta^2(\tau)}{L_1}e^{2e^2G(0)-2\pi_2/e^2L_1L_2} \label{3.14}\\ &=&
\langle \bar\psi(x)\gamma_5\psi(x)\rangle \nonumber \eeq (see \cite{Aza} and
\cite{SW}).

In the limit $L_1=L_2=L\to\infty$ from (\ref{2.29})-(\ref{2.32}) it follows
that \be
2e^2G(0)=-\ln \eta^2(\tau)+\ln\frac{m}{4\pi}+\gamma+\ln L_1, \ee where
$m=\frac{e}{\sqrt{\pi}}$ and $\gamma$ is the Euler constant, and

\be
\langle \bar\psi(x)\psi(x)\rangle
=\frac{e}{\sqrt\pi}\frac{e^\gamma}{2\pi}\label{3.16} \ee

\subsection{Correlation functions for fermion bilinears}

Using our definition of $M(x)$ we will consider the correlation functions
for gauge
invariant (regularized) bilinears

\beq
& &\langle M(x)M'(y)\rangle\nonumber\\
&&=\mathop{\lim\limits_{{\zeta\to0}\atop{\zeta'\to0}}}\left\{\langle
\bar{\psi}(x+\zeta)\tilde{\Gamma}\psi(x-\zeta)\bar{\psi}(y+\zeta')
\tilde{\Gamma}'\psi(y-\zeta')\rangle\right.\label{e316}\\ &&-
\left.\langle\bar{\psi}(x+\zeta)\tilde{\Gamma}\psi(x-\zeta)\rangle
\langle\bar{\psi}(y+\zeta')\tilde{\Gamma}'\psi(y-\zeta')\rangle\right\},
\nonumber \eeq
where $\tilde{\Gamma}=\Gamma\exp\left\{ie\int\limits_{x-\zeta}^{x+\zeta}
A_\sigma(x')dx'_\sigma\right\}$,
$\tilde{\Gamma}'=\Gamma'\exp\left\{ie\int\limits_{y-\zeta'}^{y+\zeta'}
A_\delta(y')dy'_\delta\right\}$,

$\Gamma,\Gamma'=\gamma_\mu,\gamma_5$, {\bf 1}.

{}From the general formula (\ref{q11}) we have \beq
&&\langle\bar{\psi}(x+\zeta)\tilde{\Gamma}\psi(x-\zeta)\bar{\psi}(y+\zeta')
\tilde{\Gamma}'\psi(y-\zeta')\rangle\nonumber\\
&&=\frac{1}{Z}\sum_{k}\int_{{\cal
A}_k}{\cal D}\bar{\psi}{\cal D}\psi {\cal D}Ae^{-S[A]-
S_F[\bar{\psi},\psi,A]}\label{e317}\\
&&\times\bar{\psi}(x+\zeta)\tilde{\Gamma}\psi(x- \zeta)\bar{\psi}(y+\zeta')
\tilde{\Gamma}'\psi(y-\zeta')\nonumber\\ &&=\frac{1}{Z}\sum_{k}\int_{{\cal
A}_k}{\cal
D}Ae^{-S[A]}I_{(k)}[x,\zeta; y,\zeta';\Gamma,\Gamma']\nonumber
\eeq
and only gauge sectors $C{\cal H}^{(k)}$, with $k=0,\pm1,\pm2$ contribute.

The trivial sector $k=0$. For this sector

\beq
&&I_{(0)}[x,\zeta;y,\zeta';\Gamma,\Gamma']\nonumber\\ &&=\left\{-
T_{\mu\nu}[a,b,x,\zeta,y,\zeta';\Gamma,\Gamma']S_t^{(\mu)} (x-\zeta-y-
\zeta')S_t^{(\nu)}(y-\zeta'-x-\zeta)\right.\\
&&+\left.T_\mu[a,b,x,\zeta;\Gamma]T_\nu[a,b,y,\zeta';\Gamma']S_t^{(\mu)} (-
2\zeta)S_t^{(\nu)}(-2\zeta')\right\}e^{\frac{1}{2}\Gamma_{reg}^{(0)}[A]},
\nonumber
\eeq
where
\beq
&&T_{\mu\nu}[a,b,x,\zeta,y,\zeta';\Gamma,\Gamma']\nonumber\\ &&\equiv e^{-
ie\Delta a(x)-ie\Delta a(y)}\left\{{\rm Tr}(\gamma_\mu\Gamma'
\gamma_\nu\Gamma)\cosh b_1\cosh b_2+\right.\nonumber\\ &&+{\rm
Tr}(\gamma_5\gamma_\mu\Gamma'\gamma_\nu\Gamma)\sinh b_1\cosh b_2 +{\rm
Tr}(\gamma_\mu\Gamma'\gamma_5\gamma_\nu\Gamma)\cosh b_1\sinh b_2\\
&&+\left.{\rm Tr}(\gamma_5\gamma_\mu\Gamma'\gamma_5\gamma_\nu\Gamma)
\sinh b_1\sinh b_2\right\}\nonumber
\eeq
contributes to the connected part of the corresponding diagram and
$T_\mu[a,b,x,\zeta;\Gamma]$, given in (\ref{3.4}), contributes to the
disconnected part
of this diagram.
\be \Delta a(x)\equiv a(x+\zeta)-a(x-\zeta),\ee \be \Delta a(y)\equiv
a(y+\zeta')-a(y-
\zeta'),\ee \be b_1\equiv eb(x-\zeta)-eb(y+\zeta'),\ee \be b_2\equiv
eb(y-\zeta')-
eb(x+\zeta).\ee

Sectors $k=\pm1$. For the sector $k=1$

\beq
&&I_{(1)}[x,\zeta;y,\zeta';\Gamma,\Gamma']=L_1e^{\frac{1}{2}\Gamma_{reg}^{(1)}
[A]}(\det{\cal N}_A^{(1)})^{-1}\nonumber\\
&&\times\left\{\hat{\bar{\chi}}(x+\zeta)\Gamma\hat{\chi}(x-\zeta)T_\mu
[a,b;y,\zeta',\Gamma']S_t^{(1)(\mu)}(y-\zeta',y+\zeta')\right.\nonumber\\ &&+
T_\mu[a,b;x,\zeta,\Gamma]S_t^{(1)(\mu)}(x-\zeta,x+\zeta)
\bar{\hat{\chi}}(y+\zeta')\Gamma'\hat{\chi}(y-\zeta')\\ &&-
\bar{\hat{\chi}}(y+\zeta')\Gamma e^{ie\alpha(y-\zeta')}\gamma_\mu e^{-
ie\alpha'(x+\zeta)}\Gamma'\hat{\chi}(x-\zeta)S_t^{(1)(\mu)} (y-
\zeta',x+\zeta)\nonumber\\
&&-\left.\bar{\hat{\chi}}(x+\zeta)\Gamma e^{ie\alpha(x-\zeta)}\gamma_\mu e^{-
ie\alpha'(y+\zeta')}\Gamma'\hat{\chi}(y-\zeta')S_t^{(1)(\mu)} (x-
\zeta,y+\zeta')\right\},\nonumber
\eeq
$\hat{\chi}(x)=e^{ie[a(x)-ib(x)]}\chi(x)$ and $\chi(x)$ is a zero mode in
the sector
$k=1$ given in Eq.(\ref{a134}) and a similar expression for $I_{(-
1)}[x,\zeta;y,\zeta';\Gamma,\Gamma']$ for the sector $k=-1$.

Sectors $k=\pm2$. For $k=2$ we have

\beq
&&I_{(2)}[x,\zeta;y,\zeta';\Gamma,\Gamma']=L_1^2e^{\frac{1}{2}
\Gamma_{reg}^{(2)}[A]}(\det{\cal N}_A^{(2)})^{-1}\nonumber\\
&&\times\left\{\bar{\hat{\chi}}^{(0)}(x+\zeta)\Gamma\hat{\chi}^{(0)} (x-
\zeta)\bar{\hat{\chi}}^{(1)}(y+\zeta')\Gamma'\hat{\chi}^{(1)}(y-\zeta')
\right.\nonumber\\
&&+ \bar{\hat{\chi}}^{(1)}(x+\zeta)\Gamma\hat{\chi}^{(1)}(x-\zeta)
\bar{\hat{\chi}}^{(0)}(y+\zeta')\Gamma'\hat{\chi}^{(0)}(y-\zeta')\\ &&-
\bar{\hat{\chi}}^{(0)}(x+\zeta)\Gamma\hat{\chi}^{(1)}(x-\zeta)
\bar{\hat{\chi}}^{(1)}(y+\zeta')\Gamma'\hat{\chi}^{(0)}(y-\zeta')\nonumber\\
&&-
\left.\bar{\hat{\chi}}^{(1)}(x+\zeta)\Gamma\hat{\chi}^{(0)}(x-\zeta)
\bar{\hat{\chi}}^{(0)}(y+\zeta')\Gamma'\hat{\chi}^{(1)}(y-\zeta')\right\},
\nonumber \eeq
where $\hat{\chi}^{(0)}$ and $\hat{\chi}^{(1)}$ are two independent zero
modes of
$D_A$ operator for $k=2$ and a similar expression for $I_{(-
2)}[x,\zeta;y,\zeta';\Gamma,\Gamma']$ for the sector $k=-2$.

In these formulas the vector-potential of the electromagnetic field from the
topological sector ${\cal A}_k$, is given in (\ref{a18}), \be
S_t^{(k)}(x,y)=\gamma_\mu S_t^{(k)(\mu)}(x,y) \ee is the Green's function
of the
$D_0$ operator given in (\ref{a113}) ($S_t^{(0)(\mu)}(x,y)\equiv
S_t^{(\mu)}(x-y)$).
In general, the propagator $S^{(k)}(x,y;A)$ of the fermions in the
background field
$A_\mu(x)$ from (\ref{a18})
can be expressed as follows \be
S^{(k)}(x,y;A)=e^{ie\alpha(x)}S_t^{(k)}(x,y)e^{-
ie\alpha'(y)},\ee where $\alpha(x)=a(x)-i\gamma_5b(x)$,
$\alpha'(y)=a(y)+i\gamma_5b(y)$, (see Appendix G) \subsubsection{Currents
$\Gamma=i\gamma_\alpha$, $\Gamma'=i\gamma_\beta$. ($M(x)\equiv
j_\alpha(x)$, $M(y)\equiv j_\beta(y)$)}

In this case we have contribution only from the trivial sector $k=0$. The
evaluation
of the connected and disconnected parts is straightforward (see Appendix
F): \beq
\langle j_\alpha(x)j_\beta(y)\rangle_c&=& 2\langle S_t^{(\beta)}(x-y)
S^{(\alpha)}_t(y-x)+S_t^{(\alpha)}(x-y)S_t^{(\beta)}(y-x)\nonumber\\ &-
&\delta^{\alpha\beta}S_t^{(\mu)}(x-y)S^{(\mu)}_t(y-x)\rangle^{(0)}_t, \eeq

\be
\langle j_\alpha(x)j_\beta(y)\rangle_d=
-\left\{\frac{1}{\pi^2}\frac{\zeta_\alpha}{|\zeta|^2}\frac{\zeta'_\beta}
{|\zeta'|^2}+\frac{e^2}{\pi^2}\epsilon_{\alpha\rho}\epsilon_{\beta\sigma}
\partial_\rho\partial_\sigma G(x-y)+R_{\alpha\beta}\right\}, \ee

\be
R_{11}=\frac{1}{2\pi^2}[\langle K_1(t)K_1(t)\rangle^{(0)}_t- \langle
K_2(t)K_2(t)\rangle^{(0)}_t],
\ee
\be
R_{12}=\frac{1}{\pi^2}\langle K_1(t)K_2(t)\rangle^{(0)}_t. \ee Applying an
`addition
theorem' of the $\vartheta$-functions \be \vartheta_1(z+w)\vartheta_1(z-
w)=\frac{1}{\vartheta_4^2(0)} (\vartheta_1^2(z)\vartheta_4^2(w)-
\vartheta_4^2(z)\vartheta_1^2(w)) \ee and using the `light cone components'
$j_\pm(x)=j_1(x)\mp ij_2(x)$, we get for the
`connected' part a simple result in terms of the Weierstrass function
$\wp(z)$ \cite
{Whit} \cite {Batem}

\be
\langle j_+(x)j_+(0)\rangle_c=\langle j_-(x)j_-(0)\rangle^*_c=
-\frac{1}{\pi^2L^2_1}\{\wp(z)-\langle\wp(\bar t)\rangle_t^{(0)}\}, \ee

\be
\langle j_+(x)j_-(0)\rangle_c=0,
\ee

\be
\wp(z)=-\frac{d^2}{dz^2}\ln\vartheta_1(z\mid\tau)+\frac{1}{3}
\frac{\vartheta'''_1(0\mid\tau)}{\vartheta'_1(0\mid\tau)}, \ee or returning
to Cartesian
components

\beq
\langle j_1(x)j_1(0)\rangle_c&=&-\frac{1}{4\pi^2L_1^2}\{\wp(z)+ \wp(\bar z)-
\langle\wp(\bar t)\rangle_t^{(0)}-
\langle\wp(t)\rangle_t^{(0)}\}\nonumber\\ &=&- \langle j_2(x)j_2(0)\rangle_c,
\eeq

\beq
\langle j_1(x)j_2(0)\rangle_c&=&\frac{i}{4\pi^2L_1^2}\{\wp(\bar z)
-\langle\wp(t)\rangle_t^{(0)}-\wp(z)+
\langle\wp(\bar t)\rangle_t^{(0)}\}\nonumber\\ &=&\langle
j_2(x)j_1(0)\rangle_c. \eeq

The averaging over the toron configuration results in \beq \langle\wp
(t)\rangle_t^{(0)}&=&L_1^2\langle[K_1(t)+iK_2(t)]^2\rangle^
{(0)}_t+\frac{\pi}{|\tau|}+\frac{1}{3}\frac{\vartheta'''_1(0\mid\tau)}
{\vartheta(0\mid\tau)}\nonumber\\
&=&\langle\wp (\bar t)\rangle^{(0)*}_t
\eeq
and
\be
\langle j_1(x)j_1(0)\rangle_c=\frac{1}{\pi}\partial_2^2G_0(x)+
\frac{1}{2\pi^2}[\langle
K_1(t)K_1(t)\rangle^{(0)}_t-\langle K_2(t)K_2(t) \rangle^{(0)}_t], \ee

\be
\langle j_1(x)j_2(0)\rangle_c=\frac{1}{\pi}\partial_1\partial_2G_0(x)+
\frac{1}{\pi^2}\langle K_1(t)K_2(t)\rangle^{(0)}_t. \ee Adding the
connected and
disconnected parts we get for the correlation function of
the currents
\be
\langle j_\alpha(x)j_\beta(y)\rangle=\frac{1}{\pi}\epsilon_{\alpha\rho}
\epsilon_{\beta\sigma}\partial_\rho\partial_\sigma G_m(x-y). \ee

\subsubsection{Clustering}

The clustering property means that the vacuum matrix element of a product of local
operators factorizes when their space-like separation becomes large.
Usually this
property is well established in the theories, where we have massive
particles from
the beginning. We will prove that in the Schwinger model (where massless
fermions
interact with a gauge field) this property also holds.

In order to do that we should calculate the four point function\newline
$\langle\bar{\psi}(x)\psi(x)\bar{\psi}(0)\psi(0)\rangle$, and as we have
demonstrated
in the previous section for this aim one should consider all topological
sectors with
$|k|\leq2$. In fact, in this case only the sectors $k=0$ and $|k|=2$
contribute (see
Eq.(\ref{e316})). Then the $L_1=L_2=L\to\infty$ limit should be taken, and
we will
prove that as in the case of the Schwinger model of the sphere \cite{Camil} the
trivial sector gives only a half of the value expected, the other half
comes from the
sector $|k|=2$.

{}From the general expression (\ref{e317}) by calculating traces we obtain:
\begin{eqnarray}
&&\langle\bar{\psi}(x)\psi(x)\bar{\psi}(0)\psi(0)\rangle= -2R(x,0)\langle
S_t^{(\mu)}(x)S_t^{(\mu)}(-x)\rangle_t^{(0)}\nonumber\\ &&+V(x,0)\left\{\langle
\bar{\chi}^{(0)}(x)\chi^{(0)}(x)\bar{\chi}^{(1)}(0)\chi^{(1)}(0)+
\bar{\chi}^{(1)}(x)\chi^{(1)}(x)\bar{\chi}^{(0)}(0)\chi^{(0)}(0)\right.
\nonumber\\ &&-
\bar{\chi}^{(0)}(x)\chi^{(1)}(x)\bar{\chi}^{(1)}(0)\chi^{(0)}(0)-
\bar{\chi}^{(1)}(x)\chi^{(0)}(x)\bar{\chi}^{(0)}(0)\chi^{(1)}(0)\rangle^{(2)
}_t\\
&&+\langle\bar{\phi}^{(0)}(x)\phi^{(0)}(x)\bar{\phi}^{(1)}(0)\phi^{(1)}(0)+
\bar{\phi}^{(1)}(x)\phi^{(1)}(x)\bar{\phi}^{(0)}(0)\phi^{(0)}(0)\nonumber\\ &&-
\left.\bar{\phi}^{(0)}(x)\phi^{(1)}(x)\bar{\phi}^{(1)}(0)\phi^{(0)}(0)-
\bar{\phi}^{(1)}(x)\phi^{(0)}(x)\bar{\phi}^{(0)}(0)\phi^{(1)}(0) \rangle^{(-
2)}_t\right\}\nonumber
\end{eqnarray}
where $R(x,0)=e^{4e^2[G(0)-G(x)]}$ and
$V(x,0)=\frac{\eta^2(\tau)|\tau|L_1^2}{4} e^{4e^2[G(0)+G(x)]}e^{-\frac{8\pi
}{m^2L_1L_2}}$.

One can easily show that the contributions from the sectors with $k=+2$ and
$k=-2$
are equal.

So
\begin{eqnarray}
\langle \bar{\psi}(x)\psi(x)\bar{\psi}(0)\psi(0)\rangle=-2R(x,0) \langle
S_t^{(\mu)}(x)S_t^{(\mu)}(-x)\rangle_t^{(0)} \nonumber\\
+V(x,0)\langle|\chi_1^{(0)}(x)\chi_1^{(1)}(0)-\chi_1^{(1)}(x)\chi_1^{(0)}(0)|^2
\rangle_t^{(2)},\label{3.44}
\end{eqnarray}
where $\chi^{(n)}_1(x)$ is the first component of the zero mode $\chi
^{(n)}(x)$ with
the positive chirality.

For the $t$-averaging in the trivial sector $(k=0)$ we get \be \langle
S_t^{(\mu)}(x)S_t^{\mu}(-x)\rangle_t^{(0)}=-\frac{\eta^6(\tau)}
{|L_1\vartheta_1(z\mid\tau)|^2}e^{\frac{2\pi}{L_1L_2} x_2^2}\label{3.45} \ee
and
\be R(x,0)=e^{4e^2G(0)}e^{-4\pi G_0(x)}e^{4\pi G_m(x)}.\label{3.46}\ee

{}From the general expression (\ref{a142}) for the zero modes with the positive
chirality in the sector $k=2$ we have two of them

\begin{eqnarray}
\chi_1^{(0)}(x)=
(\frac{4}{|\tau|})^{1/4}\frac{1}{L_1}e^{i\frac{e}{2}x_{\mu}t_{\mu}
+\frac{\pi}{|\tau|}(z'^2-
z'\bar{z}')}\vartheta_3(2z'|2\tau). \label{3.47} \end{eqnarray}
and

\begin{eqnarray}
\chi_1^{(1)}(x)=
(\frac{4}{|\tau|})^{1/4}\frac{1}{L_1}e^{i\frac{e}{2}x_{\mu}t_{\mu}
+\frac{\pi}{|\tau|}(z'^2-
z'\bar{z}')-\frac{\pi |\tau|}{2}+2\pi iz'} \vartheta_3(2z'+\tau |2\tau)
\end{eqnarray} Using the relation between $\vartheta$-functions \cite {Batem}
\begin{eqnarray}
\vartheta_3(z+\frac{\tau}{2}|\tau)=e^{-i\pi(z+\frac{\tau}{4})}\vartheta_2(z|
\tau),
\end{eqnarray}
$\chi_1^{(1)}(x)$ can be rewritten as follows

\begin{eqnarray}
\chi_1^{(1)}(x)=
(\frac{4}{|\tau|})^{1/4}\frac{1}{L_1}e^{i\frac{e}{2}x_{\mu}t_{\mu}
+\frac{\pi}{|\tau|}(z'^2-
z'\bar{z}')}\vartheta_2(2z'|2\tau). \end{eqnarray} With the help of the formula
\begin{eqnarray}
\vartheta_3(2v|2\tau)\vartheta_2(2u|2\tau)-
\vartheta_2(2v|2\tau)\vartheta_3(2u|2\tau)= \vartheta_1(v+u|\tau)\vartheta_1(v-
u|\tau), \end{eqnarray}
which follows from formulas (118) and (128) in \cite {Tolke}, we obtain

\begin{eqnarray}
\chi_1^{(0)}(x)\chi_1^{(1)}(0)-\chi_1^{(1)}(x)\chi_1^{(0)}(0) =
\frac{2}{|\tau|^{\frac{1}{2}}}\frac{1}{L_1^2}e^{i\frac{e}{2}x_2t_2
+\frac{\pi}{|\tau|}(z^2-
z\bar{z})+\frac{\pi}{2|\tau|}(\bar{t}^2-t\bar{t})} \nonumber\\
\times\vartheta_1(z|\tau)\vartheta_1(z+\bar{t}|\tau)e^{\frac{2\pi
x_2\tilde{t}_1} {L_1}}
\end{eqnarray}
and
\begin{eqnarray}
\langle
|\chi_1^{(0)}(x)\chi_1^{(1)}(0)-\chi_1^{(1)}(x)\chi_1^{(0)}(0)|^2\rangle_t^{
(2)}=
\frac{4}{|\tau|L_1^4}e^{\frac{4\pi}{L_1L_2}x_2^2}|\vartheta_1(z|\tau)|^2
\nonumber\\
\times\langle|\vartheta_1(z+\bar{t}|\tau)|^2e^{-2\pi
|\tau|\tilde{t}_1^2+\frac{4\pi
x_2\tilde{t}_1}{L_1}}\rangle_t^{(2)}.
\end{eqnarray}
One can easily check that in the last formula \begin{eqnarray}
\langle...\rangle_t^{(2)}=e^{\frac{2\pi x_2^2}{L_1L_2}}. \end{eqnarray} So
\begin{eqnarray}
\langle|\chi_1^{(0)}(x)\chi_1^{(1)}(0)-\chi_1^{(1)}(x)\chi_1^{(0)}(0)|^2
\rangle_t^{(2)}=
\frac{4}{|\tau|L_1^4}e^{-\frac{2\pi x_2^2}{L_1L_2}}|\vartheta_1(z|\tau)|^2.
\label{3.55}
\end{eqnarray}
{}From (\ref{3.44})-(\ref{3.46}) and (\ref{3.55}) we get
\begin{eqnarray}\langle\bar{\psi}(x)\psi(x)\bar{\psi}(0)\psi(0)\rangle
=2\frac{\eta^4(\tau)}{L_1^2}e^{4eG(0)}[e^{4\pi G_m(x)}+e^{-4\pi
G_m(x)-\frac{8\pi}
{m^2L_1L_2}}]
\end{eqnarray}
and from (\ref{2.32}) and (\ref{3.14})
\begin{eqnarray}
\langle\bar{\psi}(x)\psi(x)\bar{\psi}(0)\psi(0)\rangle
=\langle\bar{\psi}(x)\psi(x)\rangle^2\cosh4\pi \bar{G}_m(x). \label{3.57}
\end{eqnarray}
So in the flat space limit we have, using (\ref{3.16}) \be
\langle\bar{\psi}(x)\psi(x)\bar{\psi}(0)\psi(0)\rangle{\mathop{=}\limits_{L_
1=L_2=L\to
\infty}} \frac{e^{2\gamma}}{\pi}\frac{e^2}{8\pi^2}\left\{e^{2K_0(m|x|)} +e^{-
2K_0(m|x|)}\right\}
\ee
and when $|x|\to\infty$ with the help of (\ref{2.33}) \begin{eqnarray}
\langle\bar{\psi}(x)\psi(x)\bar{\psi}(0)\psi(0)\rangle\to_{|x|\to\infty}
\left(\frac{e}{\sqrt{\pi}}\frac{e^\gamma}{2\pi}\right)^2=\langle\bar{\psi}
\psi \rangle^2.
\end{eqnarray}
Thus the clustering property is indeed satisfisfied.

\newpage

{\bf Conclusions}
\vspace{1cm}

Using the path integral approach we have performed a complete 
analysis of the
Schwin-\
ger model on the torus. This model is not just a 
generalization to the
ordinary SM in the infinite space time. The compactification enables to
investigate
the model in a mathematically more satisfactory way, keeping all
singularities under
control. The relevant differential operators have discrete spectra 
and the path
integrals in the quantum theory can be properly defined if one uses their
eigenfunctions.

All configurations of the electromagnetic potential (abelian gauge field)
defined on
the torus can be classified according to their topological charge. The
nontrivial
gauge field topology implies the occurrence of fermionic zero modes, which
need a
special treatment in the quantum theory and contribute to correlation
functions of
fermionic fields. For gauge fields of the topological charge $k$, the
massless Dirac
operator possesses exactly $|k|$ zero modes. We have obtained explicit
expressions for them and calculated the spectrum of the Dirac operator in any
topological sector.

In quantum theory using the Pauli-Villars regularization to remove the
ultraviolet
divergences occurring in the fermionic path integral, we have calculated
the part of
the gauge field effective action, which appears due to the fermion integration
(Eq.(\ref{2.24})). Effect of the torons (zero modes of the gauge potential)
on the fermion
fields reveals itself in the toron effective action $\Gamma^{(0)}(t)$ (Eq.
(\ref{3.12})),
which rules their dynamics and controls infrared singularities.

We have found the propagator of the gauge field on the torus, which is
expressed in
terms of massless and massive Green's functions of the Laplacian on the
torus, and
calculated the fermionic propagators in the background toron field and in
topologically nontrivial gauge field (Appendices F and G).

Finally, we have explicitly calculated several expectation values of
physical interest.
Toron averaging (Eq.(\ref{3.11})) assures a translational invariant
distribution of the
symmetry breaking zero modes in the topologically non-trivial sectors. For the
two-and four-point functions of fermion fields sectors with $|k|\leq1$ and
$|k|\leq2$,
respectively, contribute. For two point function the correct result (found
before
by operator methods) is obtained only, if the presence of the zero modes are
properly accounted for and their role in the chiral symmetry breaking by an
anomaly
becomes particularly transparent. We have calculated the four-point function
$\langle\bar{\psi}\psi(x)\bar{\psi}\psi(0)\rangle$ and proved the
clustering property.

Comparing the torus compactification of the geometric \cite {JA} with the
ordinary
SM, which is considered in the present work, one can find many
similiarities. But
there are also some differences. The effective action in the geometric SM has a
factor 2, which implies a factor $\sqrt 2$ in the mass of the isoscalar
particle (in the
geometric SM we have additional internal symmetry and isospin multiplet of
massles particles). The factor 2 in the toron action changes the character
of the
integration over the torons considerably, and hence the dynamical role of
the torons
in these cases.

Nowdays there are very intensive discussions in the literature on finite
temperature
(size) effects in quantum field-theoretical models. To study them in the present
context one should let the spacial (`temporal') extension of $L_2(L_1)$ tend to
infinity.

It would be interesting to extend this investigation to the cases of the chiral
Schwinger, massive Schwinger and Thirring models.

These problems remain to be considered in the future.

\vskip2truecm

{\bf Note added}
After finalizing this work I became aware of a recent paper by Fayyazuddin et al
{\cite {Fay}, where the same result for the four point fermionic function (Eq.
(\ref{3.57})) was obtained. I thank A.Wipf for bringing this reference to
my attention.

%\newpage
\vspace{2cm}
{\bf Acknowledgments}
\vspace{0.7cm}

I am very grateful to Prof. H.Joos (DESY, Hamburg) who initiated this
investigation
and has been patiently sharing with me the experience and knowledge he gained
working with the Schwinger Model for several years. Many results presented
in this
work were obtained in collaboration with him. This was done during the
realization of
our joint projects, when I several times visited DESY, which I would like
to thank for
hospitality.

Discussions with A.Wipf and I.Sachs at ETH (Z\"{u}rich), F.Strocchi and
G.Morchio
(Scuola Normale Superiore, Pisa) and P.Marchetti at the University of
Padova were
extremely useful.

I am very thankful to IASBS (Zanjan, Iran) and ICTP (Trieste, Italy), where
this work
was finished, for providing excellent conditions to carry out effective
research and a
stimulating atmosphere.

\vspace{1cm}
\set
\appendix {\bf Appendix A. Derivation of the Eq.(\ref{q2}).} \vspace{0.7cm}
\renewcommand{\theequation}{A.\arabic{equation}}

As it has been mentioned the nonzero eigenvalues of the $D$ operator appear in
pairs: $i E_{\nu}$ and $-iE_{\nu}(E_{\nu}$ is real). Therefore \begin{equation}
{\det}'D=\prod _{\nu}(iE_{\nu})(-iE_{\nu})=\prod _{\nu} E_{\nu}^2
\end{equation} and
\begin{equation}
{\det}'D^{\dagger}=\prod_{\nu}(-iE_{\nu})(iE_{\nu})=\prod_{\nu}E_{\nu}^2.
\end{equation}
Thus
\begin{equation}
{\det}'D=\sqrt {{\det}'DD^{\dagger}}. \end{equation} Furthermore
\begin{eqnarray}
\det(D-M_iL_1)&=&\prod
_{\nu}(-iE_{\nu}-M_iL_1)(iE_{\nu}-M_iL_1)(-M_iL_1)^{| k|} \nonumber \\
&=&\prod _{\nu}(E_{\nu}^2+M_i^2L_1^2)(-M_iL_1)^{|k|}, \end{eqnarray} since
in the
sector with Pontriyagin index $k$ the $D$ operator has $|k|$ zero modes. \\
On the other hand
\begin{eqnarray}
\det (DD^{\dagger}+M_i^2L_1^2)&=&\det[(D-M_iL_1)(-D-M_iL_1)] \nonumber \\
&=&\det(D-M_iL_1)\det(-D-M_iL_1) \\
&=& \prod_{\nu}(E_{\nu}^2+M_i^2L_i^2)^2(-M_iL_1)^ {2|k|}.\nonumber
\end{eqnarray}
So
\begin{equation}
\det(D-M_iL_1)=(-1)^ k \det(DD^{\dagger}+M_i^2L_1^2)^{1/2} \end{equation} and
from (\ref{q1})
\begin{equation}
\exp \frac {1}{2}
\Gamma_{reg}^{(k)}[A]=(-1)^{|k|}({\det}'DD^{\dagger})^{1/2} \prod _{i=1}^{r}
\det(DD^{\dagger} +M_i^2L_1^2)^{\frac {e_i}{2}}. \end{equation}

\vspace{1cm}
\set
\appendix {\bf Appendix B. Derivation of the Eq.(\ref{q4}).} \vspace{0.7cm}
\renewcommand{\theequation}{B.\arabic{equation}}

First, let us prove that for any operator $\hat{A}$ \begin{equation} \delta
{\rm Tr}(e^{-
t\hat{A}})=-t{\rm Tr}(\delta \hat{A}e^{-t\hat{A}}). \end{equation}
We have
\begin{eqnarray}
\delta e^{-t\hat{A}}&=&e^{-t(\hat{A}+\delta
\hat{A})}-e^{-t\hat{A}}=e^{-t\hat{A}} e^{-
t\delta \hat{A}}e^{t[\hat{A},\delta \hat{A}]}-e^{-t\hat{A}} \nonumber \\
&&+O((\delta
A)^2)=e^{-t\hat {A}}(-t\delta \hat {A}+t[\hat{A},\delta \hat{A}]
+O((\delta\hat{A})^2))
\end{eqnarray}
Since ${\rm Tr}$ is a linear operation
\begin{equation}
\delta {\rm Tr}(e^{-t\hat{A}})=
{\rm Tr}(\delta e^{-t\hat{A}})=-t {\rm Tr}(\delta\hat{A}e^{-t\hat{A}}).
\label{b3}
\end{equation}
Second, let us find the variation $\delta D$ under the variation of $b(x)$

\[ D=L_1\gamma_{\mu}(\partial_{\mu}-ieA_{\mu})=L_1\gamma _{\mu}\partial _{\mu}
+eL_1\gamma_{\mu}\gamma_5\partial_{\mu}b(x)+\cdots, \] where we have not
written the terms which do not vary under the variation of $b(x)$. For any
function
$f(x)$ we have
\begin{eqnarray}
(\delta Df)(x)&=&(D(b+\delta b)f)(x)-(Df)(x)\nonumber \\ &=&eL_1\gamma _{\mu}
\gamma_5(\partial_{\mu}b(x)+\partial_{\mu}\delta b(x)) f(x)-eL_1\gamma
_{\mu}\gamma_5 \partial_{\mu}b(x)f(x) \nonumber \\ &=&eL_1\gamma
_{\mu}\gamma_5(\partial_{\mu}\delta b(x))f(x)\nonumber \\ &=&eL_1\gamma
_{\mu}\gamma_5[\partial_{\mu},\delta b(x)]f(x)=-e\gamma_5 [D,\delta b(x)]f(x),
\end{eqnarray}
since
\[ \gamma _{\mu}\gamma_5=-\gamma_5\gamma _{\mu}. \] Thus \begin{equation}
\delta D=-e\gamma_5[D,\delta b(x)].
\end{equation}
Now let us find the variation $\delta (DD^{\dagger})$: \begin{eqnarray} \delta
(DD^{\dagger})&=&-\delta (D^2)=-D\delta D-(\delta D)D \nonumber \\
&=&e[D\gamma_5D\delta b-D\gamma_5\delta b D+\gamma_5D\delta bD-
\gamma_5 \delta b DD] \nonumber \\
&=&e\gamma_5[DD^\dagger\delta b+\delta b DD^\dagger+2D\delta bD],
\end{eqnarray}
since
\begin{equation}
D\gamma_5=-\gamma_5 D=\gamma_5D^{\dagger}.\label{b8}\end{equation}

Finally from (\ref{b3}) and (\ref{b8}) we get \begin{eqnarray} \delta ({\rm
Tr}e^{-
tDD^{\dagger}})&=&-t{\rm Tr} (\delta (DD^{\dagger})e^{- tDD^{\dagger}})
\nonumber
\\
&=&-4et {\rm Tr}(\gamma_5\delta b DD^{\dagger}e^{-tDD^{\dagger}}) =4et\frac
{d}{dt}
(\gamma_5\delta be^{-tDD^{\dagger}}).
\end{eqnarray}

\newpage
\set
\appendix {\bf Appendix C. L\"{u}sher's formula.} \vspace{0.7cm}
\renewcommand{\theequation}{C.\arabic{equation}}

We can write the following representation

\beq
&&{\rm Tr}'{\sum _{i=1}^{r}e_i\ln(D_0D_0^{\dagger}+M_i^2L_1^2)}\nonumber\\
&&=\sum_{i=1}^{r}e_i\int_{0}^{\infty}\frac{dt}{t}{\rm
Tr}'{e^{-t(D_0D_0^\dagger)}} e^{-
tM_i^2L_1^2}\label{c1}
\eeq
{}From (\ref{q9}) we get
\begin{equation}
\zeta'(s\mid D_0D_0^{\dagger})=-\sum _{m}(\ln\lambda_m)d_m \lambda_m^{-s}.
\end{equation}
So
\begin{equation}
\zeta '(0\mid D_0D_0^{\dagger})=-\ln\prod_{m}\lambda_m^{d_m}=-\ln {\det}'
(D_0D_0^{\dagger})=-{\rm Tr}'\ln(D_0D_0^{\dagger}). \end{equation} Again from
(\ref{q9}) we have
\begin{equation}
\zeta (s\mid D_0D_0^{\dagger} )=\sum_{m}d_m \frac
{1}{\Gamma(s)}\int_{0}^{\infty}\frac {dt} {t}t^se^{-t
\lambda_m}=\frac{1}{\Gamma(s)}\int_{0}^{\infty}\frac{dt}{t} t^s{\rm Tr}'(e^{-
t(D_0D_0^{\dagger})}).
\end{equation}
Thus we see that $\Gamma(s)\zeta (s\mid D_0D_0^{\dagger})$ is a Mellin
transformation of the ${\rm Tr}'(e^{-t(D_0D_0^{\dagger})})$, so we can find
${\rm
Tr}'(e^{-t(D_0D_0^{\dagger})})$ doing the inverse transformation of
\begin{equation}
\Gamma (s)\zeta (s\mid D_0D_0^{\dagger})=\frac {r_1(D_0D_0^{\dagger})\Gamma
(s)}{s-1} +A\Gamma (s)+\Gamma (s)\varphi(s), \label{c5} \end{equation} where we
have used (\ref{q12}). $A$ is a constant: $A=\zeta(0\mid D_0D_0^ {\dagger})-
r_1(D_0D_0^{\dagger})$ and $\varphi (s)$ is an entire function of $s$
and
$\varphi (0)=0.$
\begin{equation}
{\rm Tr}'(e^{-tD_0D_0^{\dagger} })=\frac {1}{2\pi i} \int_{\sigma_1-i
\infty}^{\sigma_1
+i\infty}\Gamma (s)\zeta (s\mid D_0D_0^{\dagger})t^{-s}ds,\label{c6}
\end{equation}
where $\sigma_1$ is any real number greater than unity. The second line of
(\ref{q8}) can be expressed as a sum of the two terms \begin{eqnarray} &&\sum
_{i=1}^{r}e_i\int_{0}^{\infty}\frac {dt}{t}{\rm Tr}'\{e^{-t(D_0
D_0^{\dagger})}\}e^{-
tM_i^2L_1^2} \nonumber\\ &&=\sum _{i=1}^{r}e_i\int _{0}^{1}\frac
{dt}{t}{\rm Tr}'\{e^{-
t(D_0D_0^{\dagger})}\} e^{-tM_i^2L_1^2}+\sum
_{i=1}^{r}e_i\int _{1}^{\infty}\frac {dt}{t}{\rm Tr}'\{
e^{-t(D_0D_0^{\dagger})}\}e^{-
tM_i^2L_1^2}. \label{c7} \end{eqnarray}
The second term disappears in the limit $M_i\to\infty $ and only the first
term gives
ultraviolet divergencies when $M_i\to\infty $. So we should calculate the
integral in
(\ref{c7}) only for $0<t<1$. In this case we can close the contour in
(\ref{c6}) on the
left, take for the integrand its expression (\ref{c5}) and use the residuum
formula.

Then
\begin{equation}
{\rm Tr}'(e^{-tD_0D_0^{\dagger}})=\frac{r_1(D_0D_0^{\dagger})}{t}-r_1
(D_0D_0^{\dagger})+A+\cdots,
\end{equation}
where $\cdots$ is a sum of the positive powers of $t$ which also do not
contribute to
the integral in the limit $M_{i}\to\infty$. So for the first sum in the
second line of
(\ref{q8}) we have \begin{equation}
\sum_{i=1}^{r}e_i
\int_{0}^{1}\frac{dt}{t}\left(\frac{r_1(D_0D_0^{\dagger})} {t}+\zeta (0 \mid
D_0D_0^{\dagger})\right)e^{-tM_i^2L_1^2}\label{c9} \end{equation} up to terms,
which disappear in the limit $M_i\to\infty $. The expression in (\ref{c9})
is equal to
\begin{equation}
\sum_{i=1}^r e_i(-r_1(D_0 D_0^{\dagger}) M_i^2 L_1^2 \ln M_i^2 L_1^2 +\zeta
(0\mid D_0D_0^{\dagger})\ln M_i^2 L_1^2).\label{c10} \end{equation} So from
(\ref{c1}) and (\ref{c7})--(\ref{c10}), we obtain (\ref{q10}).

\set
\vspace{1cm}
\appendix {\bf Appendix D. Calculation of $r_1(D_0D_0^{\dagger})$, $\zeta (0\mid
D_0D_0^{\dagger})$ and $\zeta'(0\mid D_0D_0^{\dagger})$.} \vspace{0.7cm}

\renewcommand{\theequation}{D.\arabic{equation}} \begin{enumerate} \item{\it
Trivial sector $(k=0)$.}
\setcounter{footnote}{0}
In this case the background electromagnetic field is represented only by
the toron
field. The spectrum $\lambda_{n_1,n_2}$ of the $D_0D_0^{\dagger}$ operator%
\footnote{We should remind the reader that now we are considering dimensionless
operators $D_0$ and $D_0^{\dagger}$ obtained by the multiplication of $L_1$. }
with 2-fold degeneracy is known (see (1.47)) and \begin{eqnarray} \zeta (s\mid
D_0D_0^{\dagger})=2(4\pi ^2)^{-s}\sum _{n_1,n_2} [(n_1-\tilde{t}_1)^2+
\frac {1}{|\tau|^2}(n_2-\tilde{t}_2)^2]^{-s}\nonumber\\ =2(4\pi ^2)^{-s} Z
\left|
\begin{array}{cc}
-\tilde {t}_1&-\tilde {t}_2 \\
0&0
\end{array} \right|_{\varphi}^{s}\label{d1}, \end{eqnarray} where
$Z\left|\;\;\;\;\;\;\;
\right|_\varphi^{(s)}$ is Epstein $\zeta$ -function (see
\cite{Batem} (17.93), in our case $p=2,$ $\varphi (x)=x_1^2+\frac {1}{|
\tau|^2}x_2^2,
\;\; g_1=-\tilde {t}_1, \;\; g_2=-\tilde{t}_2, \;\; n=0)$.

This function has a pole at $s=1$ with the residue $\bar{r}_1=\pi \Delta^ {-1/2}
\Gamma (2)$, where $\Delta =\det A$ and $A$ is a matrix of the quadratic form
$\varphi (x) =\sum_{i,j} a_{ij}x_ix_j,\; A=\parallel a_{ij}\parallel$. In
our case
$\Delta =\frac {1}{|\tau|^2}$, so $\bar {r}_1=\pi|\tau|$. Thus \begin{equation}
r(D_0D_0^{\dagger})=2\frac {\pi|\tau|}{4\pi ^2}=\frac {|\tau|} {2\pi}.
\end{equation}
In order to find $\zeta (0\mid D_0D_0^{\dagger})$ and $\zeta'(0\mid D_0D_0^
{\dagger})$, i.e. $ Z(0)$ and $Z'(0)$, we will use the following method%
\footnote{This method was suggested by A.Coste.} \begin{equation}
Z(s)=\sum_{n_1,n_2}\left(\tilde{n}_1^2+\frac {1}{|\tau|^2}\tilde
{n}_2^2\right) ^{-
s}=\frac{1}{\Gamma (s)}\sum _{n_1,n_2}\int _{0}^{\infty}\frac {dt}{t}t^se^
{-(\tilde
{n}_1^2+\frac {1}{|\tau|^2}\tilde{n}_2^2)t}\equiv\frac{1} {\Gamma
(s)}J(s),\label{d3}
\end{equation}
where $\tilde{n}_i\equiv n_i-\tilde{t}_i, \;\; i=1,2$. \\ Using the Poisson
summation
formula
\begin{equation}
\sum _{n}f(n)=\sum _{p} \int _{-\infty}^{\infty}dxf(x)e^{-2\pi ipx},
\end{equation}
we have
\begin{equation}
J(s)=\sum _{p} \sum _{n_2}\int _{0}^{\infty}\frac {dt}{t}t^s
\int_{-\infty}^ {\infty}dx e^{-
2\pi ipx-[(x-\tilde{t}_1)^2+\frac {1}{|\tau|^2} \tilde{n}_2^2]t} \end{equation}
Doing the Gaussian integration over $x$ we get \begin{equation}
J(s)=\sqrt{\pi}\sum_{p}\sum_{n_2}\int_{0}^{\infty}\frac{dt}{t}\frac{t^s}
{\sqrt{t}}e^{-
[t(\frac{\tilde{n}_2^2}{|\tau|^2}+\tilde{t}_1^2)+\frac {\pi^2}{t}(p+\frac
{i\tilde{t}_1t}{\pi})^2]} \equiv \sum _{p}J_p(s). \end{equation} For the
calculation of
the value at $s=0$ we will consider two cases separately

\( a) p=0 \)

\begin{eqnarray}
J_0(s)&=&\sqrt {\pi}\sum_{n_2} \int_{0}^{\infty} \frac {dt}{t}\frac
{t^s}{\sqrt {t}}e^{-t\frac
{\tilde{n}_2^2}{|\tau|^2}}=\sqrt{\pi}\sum_{n_2} \int_{0} ^{\infty}dt
t^{s-3/2}e^{-t(\frac
{\tilde {n}_2}{|\tau|})^2} \nonumber \\
&=&\sqrt{\pi} \Gamma\left(s-\frac {1}{2}\right)|\tau|^{2s-1}\sum
_{n_2}\left| n_2-\tilde
{t}_2\right|^{-2s+1} \\
&=&\sqrt {\pi}\Gamma \left(s-\frac {1}{2}\right)|\tau|^{2s-1}
\left\{\zeta_R (2s-1,\tilde
{t}_2)+\zeta _R(2s-1,1-\tilde {t}_2)\right\}, \nonumber \end{eqnarray}
since $0<\tilde
{t}_2<1.\;\;\zeta _R(s\mid q)$ is a generalized Riemann's $\zeta$-
function
\begin{equation}
\zeta _R(s\mid q)=\sum _{n=0}^{\infty}\frac {1}{(q+n)^s}=\frac {1}{\Gamma
(s)}\int_
{0}^{\infty}\frac {t^{s-1}e^{-qt}}{1-e^{-t}}dt, \;\;\;\;\;\;\; {\rm Re}s>0.
\end{equation}
If $n$ is nonnegative integer,
\begin{equation}
\zeta _R(-n\mid q)=-\frac {B_{n+1}(q)}{n+1}, \end{equation} therefore
\begin{equation}
J_0(0)=-\frac{2\pi}{|\tau|}\left\{-\frac {B_2(\tilde {t}_2)}{2}- \frac
{B_2(1-\tilde
{t}_2)}{2}\right\}=\frac {2\pi }{|\tau|}B_2 (\tilde {t}_2)=\frac
{2\pi}{|\tau|}\left(\tilde{t}_2^2-
\tilde {t}_2+ \frac {1}{6}\right).\label{d10} \end{equation}

\( b) p\neq 0 \)

\begin{equation}
J_p(s)=\sqrt{\pi}e^{-2\pi ip\tilde{t}_1}\sum _{n_2}\int_{0}^{\infty}\frac
{dt}{t}\frac
{t^s}{\sqrt{t}}e^{-\frac {\tilde{n}_2^2}{|\tau|^2}t- \frac {\pi ^2}{t}p^2}.
\end{equation}
Using the new variables $u=\frac{|\tilde{n}_2|}{\pi|\tau|\;| p|}t, \;\; k
=\frac{|\tilde{n}_2|}{|\tau|}|p|$ and taking into account that \begin{equation}
\int _{0}^{\infty}du u^{-3/2}e^{-\pi k (u+\frac {1}{u})}=\frac {1}{\sqrt {
k }}e^{-2\pi k },
\end{equation}
we obtain for the special case $s=0$
\begin{equation}
J_p(0)=e^{-2\pi ip\tilde{t}_1}\sum_{n_2}\frac{1}{|p|}e^{-2\pi
\frac{|\tilde{n}_2|}{|\tau|}|p|}.
\end{equation}
Thus
\begin{eqnarray}
&&J(0)=J_0(0)+\sum_{p\neq 0}J_p(0)=J_0(0)\nonumber \\ &&-\sum_{n_2}\ln
\left\{(1-e^{-2\pi
\frac{|\tilde{n}_2|}{|\tau|}+i\tilde{t}_1})(1-e^{-2\pi\frac
{|\tilde{n}_2|}{|\tau|}-i\tilde
{t}_1})\right\}\nonumber \\ &&=J_0(0)-\sum_{n=1}^{\infty}\ln\left\{(1-e^{-\frac
{2\pi}{|\tau|} (n-\tilde{t}_2+i|\tau|\tilde{t}_1})(1-e^{-\frac {2\pi}{|\tau
|}(n+\tilde{t}_2-
i|\tau|\tilde{t}_1)})\right\}\nonumber \\
&&-\ln(1-e^{-\frac{2\pi}{|\tau|}(\tilde{t}_2+i|\tau|
\tilde{t}_1)})
-\sum_{n=1}^{\infty}\ln \left\{(1-e^{-\frac {2\pi}{|\tau|}(n-
\tilde{t}_2+i|\tau|\tilde{t}_1)})(1-e^{-\frac {2\pi}{|\tau |}(n+\tilde{t}_2-
i|\tau|\tilde{t}_1)})\right\} \nonumber \\
&&-\ln(1-e^{-\frac{2\pi}{|\tau|}(\tilde{t}_2-i|\tau|
\tilde{t}_1})
=J_0(0)-\ln \left\{\prod_{n=1}^{\infty}(1-e^{-\frac{2\pi}{|\tau
|}(n-\bar{t})})(1-e^{-
\frac{2\pi}{|\tau|}(n+\bar{t})})\right\} \nonumber\\ &&-\ln
\left\{\prod_{n=1}^{\infty}(1-e^{-
\frac{2\pi}{|\tau|} (n-t)})(1-e^{-\frac{2\pi}{|\tau|}(n+t)}) \right\} -\ln
\left\{ (1-e^{-
\frac{2\pi}{|\tau|}t})(1-e^{-\frac {2\pi}{|\tau|}\bar{t}})
\right\}\nonumber \\ &&=J_0(0)-
\ln\left\{(1-e^{-\frac
{2\pi}{|\tau|}t})(1-e^{-\frac {2\pi}{|\tau|}\bar{t}}) \right\}\nonumber\\
&&-\ln\left\{
\prod_{n=1}^{\infty}\left(1-2e^{-\frac{2\pi}{|\tau|}n}\cosh \frac
{2\pi}{|\tau|}\bar{t}+e^{-
\frac {4\pi}{|\tau|}n}\right)\right\}\nonumber\\ &&-\ln\left\{
\prod_{n=1}^{\infty}\left(1-2e^{-\frac{2\pi}{|\tau|}n}
\cosh\frac{2\pi}{|\tau|}t+e^{-\frac
{4\pi}{|\tau|}n}\right) \right\}.\nonumber \end{eqnarray} Now let us
recollect the
definition of the $\vartheta_1$ Jacobi's $\vartheta$ function
\cite{Batem}
\[ \vartheta _1(t\mid q)=2q_0q^{1/4}\sin \pi
t\prod_{n=1}^{\infty}(1-2q^{2n}\cos 2\pi
t+q^{4n}),\]
where $q_0=\prod_{n=1}^{\infty}(1-q^{2n})$. \par In our case we will take $q=e^{-
\frac {\pi}{|\tau|}}=e^{-i\frac {\pi} {\tau}}$. Then
\begin{eqnarray}
J(0)&=&J_0(0)-\ln\left|\vartheta_1\left(i\frac{t}{|\tau|}\left|-\frac{1}
{\tau}\right.
\right)\right|^2 +2\ln(2q_0e^{-\frac {\pi}{4|\tau|}})\nonumber \\
&+&\ln\left(\sinh \frac
{\pi}{|\tau|}t\; \sinh \frac {\pi}{|\tau|}\bar{t}\right) -\ln
\left\{(1-e^{-\frac {2\pi}{|\tau|}t})(1-
e^{-\frac{2\pi} {|\tau|}\bar{t}})\right\}, \end{eqnarray}
where $t=\tilde {t}_2+i|\tau|\tilde{t}_1. $ For the last term we have $$
\ln\left\{(1-e^{-
\frac{2\pi}{|\tau|}t})(1-e^{-\frac{2\pi}{|\tau|}\bar{t}})
\right\}=\ln\left\{e^{-
\frac{\pi}{|\tau|}t}(e^{\frac{\pi}{|\tau|}t} -e^{-\frac{\pi}{|\tau|}t})e^{-
\frac{\pi}{|\tau|}\bar{t}}\right.	$$

$$ \left.
\times (e^{\frac {\pi}{|\tau|}\bar{t}}-e^{-\frac{\pi}{|\tau|}
\bar{t}})\right \} =\ln\left\{4e^{-
\frac{\pi}{|\tau|}(t+\bar{t})}\right\}+\ln\left\{
\sinh\frac{\pi}{|\tau|}t\sinh\frac
{\pi}{|\tau|}\bar{t} \right\}. $$
So
\begin{equation}
J(0)=\frac{2\pi}{|\tau|}\left(\tilde{t}_2^2-\tilde{t}_2+\frac {1}{6}\right)
-\ln\left|\vartheta_1\left(i\frac{t}{\tau}\left|-\frac{1}{\tau}\right.\right
)\right|^2+\ln q_0^2-
\frac {\pi}{2|\tau|}-\ln e^{-\frac {2\pi}{|\tau|}\tilde{t}_2}
\end{equation}and using the
relation
\[ \vartheta_1\left(\frac {z}{|\tau|}\left| -\frac {1}{\tau}\right.\right)
=i\sqrt{|\tau|}e^{-\frac
{\pi z^2}{|\tau|}}\vartheta _1 (iz\mid \tau ), \] we finally get
\begin{eqnarray}
J(0)&=&\frac {2\pi}{|\tau|}\left(\tilde{t}_2^2-\frac
{1}{12}\right)-\ln|\tau |-\frac
{\pi}{|\tau|}(t^2+\bar{t}^2)\nonumber \\ &-&\ln\left|\vartheta_1(-t|\tau
|)\right|^2+\ln
q_0^2\left(\frac{1}{\tau}\right)\nonumber \\ &=&\frac
{2\pi}{|\tau|}\left(\bar{t}_1^2-\frac
{1}{12}\right)-\ln\left|\vartheta _1 (t\mid \tau
)\right|^2+\ln q_0^2\left(-\frac {1}{\tau}\right)-\ln|\tau|, \end{eqnarray}
where
\[ \bar{t}_1 =|\tau|\tilde{t}_1=\frac {eL_2t_1}{2\pi}. \] {}From (\ref{d1})
and (\ref{d3}) we
have \begin{equation}
\zeta (0\mid D_0D_0^{\dagger})=2\frac {1}{\Gamma (0)}J(0)=0, \;\;\;\;{\rm since}
\;\;\;\frac {1}{\Gamma (0)}=0
\end{equation}
and
\begin{eqnarray}
&&\zeta'(0\mid D_0D_0^{\dagger})=2J(0)\nonumber\\ &&=\frac
{4\pi}{|\tau|}\left(\bar
{t}_1^2 -\frac {1}{12}\right)-2\ln\left|\vartheta_1(t\mid \tau
)\right|^2+2\ln q_0^2 \left(-
\frac{1}{\tau}\right)-2\ln|\tau| \\ &&= \frac{4\pi}{|\tau|}\bar{t}_1^2-
2\ln|\vartheta_1(t\mid\tau) \eta^{-1}(\tau)|^2, \end{eqnarray}
where $\eta(\tau)$ is Dedekind's function.

\item{\it Nontrivial sector $(k\neq 0)$ }

In this case the spectrum of the $D_0D_0^{\dagger} $ operator: $\lambda
_m=\frac{4\pi|k|}{|\tau|}n$ with $2|k|$ fold degeneracy and \begin{equation}
\zeta (s\mid D_0D_0^{\dagger})=\frac {2|k||\tau|^s}
{(4\pi|k|)^s}\sum_{n=1}^{\infty}\frac {1}{n^s}=\frac {2|
k||\tau|^s}{(4\pi|k|)^s}\zeta_R(s),\end{equation} where $\zeta _R(s)$ is
Riemann's
$\zeta$ function which has a pole at $s=1$ with the residue 1 and \[
\zeta_R(0)=-
\frac{1}{2}, \;\; \zeta'_R(0)=-\frac{1}{2}\ln 2\pi. \] Thus \begin{equation}
\zeta (0\mid D_0D_0^{\dagger})=-|k|, \;\; r(D_0D_0^{\dagger})=\frac {|
\tau|}{2\pi},\end{equation}
\begin{equation}
\zeta '(s\mid D_0D_0^{\dagger})=2|k|\left\{-\left(\ln\frac {4\pi|
k|}{|\tau|}\right)\left(\frac
{|\tau|}{4\pi|k|}\right)^s \zeta _R(s)+\frac {|\tau|^s}{(4\pi|k|)^s}\zeta
'_R(s) \right\},
\end{equation} \begin{equation}
\zeta'(0\mid D_0D_0^{\dagger})=-|k|\ln\frac {4\pi|k|}{|\tau|}+|k|\ln 2\pi =-
|k|\ln\frac{2|k|}{|\tau|}.\end{equation} \end{enumerate} \vspace{1cm}
\set
\appendix {\bf Appendix E. The Green's function of the Laplacian on the Torus.}
\vspace{0.7cm}
\renewcommand{\theequation}{E.\arabic{equation}}

For the function $G_0(x)$ we have the following equation \begin{equation} -\Box
G_0(x)=\delta (x)-\frac {1}{L_1L_2}, \label{e1} \end{equation} where
$\delta(x)$ is a
$\delta$-function on the torus \begin{equation} \delta (x)=\frac
{1}{L_1L_2}\sum
_{n_1,n_2}e^{\frac {2\pi}{L_1}(n_1x_1+| \tau|^{-
1}n_2x_2)}.
\end{equation}
Let us introduce the function
\begin{equation}
\tilde{G} (x)=G_0(x)-x^2/4L_1L_2.\label{e.3}\end{equation} Then from
(\ref{e1}) it
follows that
\begin{equation}
-\Box \tilde {G}(x)=\delta (x)
\end{equation}
Since the $G_0(x)$ function obeys the periodicity condition \begin{equation}
G_0(x+\hat{\nu}L_{\nu})=G_0(x),\label{e5} \end{equation} for the
$\tilde{G}(x)$ we
have
\begin{equation}
\tilde{G}(x+\hat{\nu}L_{\nu})=G_0(x)-\frac {(x+\hat{\nu}L_{\nu})^2}{4L_1L_2}=
\tilde{G}(x)-\frac {x_{\nu}L_{\nu}}{2L_1L_2}-\frac {L_{\nu}^2}{4L_1L_2}.
\end{equation}
For the $\tilde{G}(x)$ function we will choose the following ansatz
\begin{equation}
\tilde{G}(x)=-\frac {1}{2\pi}{\rm Re}\ln\sigma (z), \end{equation} where
$z$ is defined
in (1.14), and $\sigma (z)$ is analytic in the box $(1,\tau)$ and
has a single zero at $z=0$ \begin{equation} \sigma (z)=z+\cdots. \label{e8}
\end{equation}
So
\begin{equation}
\tilde{G}(x)=-\frac{1}{2\pi}\ln|z|+\cdots ,\;\;\; {\rm when}
\;\;\;|z|\rightarrow 0.
\end{equation}
Taking into account the behaviour of the $\sigma (z)$ function for small
$|z|$ and
the periodicity condition (\ref{e5}), we choose it in the following form
\begin{equation}
\sigma (z)=ce^{az^2+bz}\vartheta_1(z\mid \tau). \end{equation} Since the
$\vartheta_1(z\mid \tau)$ function obeys the following periodicity conditions
$$\vartheta _1(z+1\mid\tau )=-\vartheta_1(z\mid\tau),$$ $$\vartheta
_1(z+1\mid\tau)=-e^{-i\pi(2z+\tau)}\vartheta_1(z\mid\tau),$$ we get the
following
equations for the constants $a$ and $b$

$$ \ln|\sigma (z+1)|={\rm Re}(2az+a)+{\rm Re}\;
b+\ln\left|\sigma\left(\frac{z}{L_1}
\right)\right|=\pi \frac {x_1}{L_2}+\frac {\pi}{2|\tau|}+\ln|\sigma (z)|,$$
\begin{eqnarray}
\ln|\sigma (z+\tau)|&=&{\rm Re}(2aiz|\tau|-a|\tau|^2)- {\rm Im} b+2\pi \frac
{x_2}{L_1}+\pi|\tau|+\ln|\sigma(z)|\nonumber \\
&=&\pi\frac{x_2}{L_1}+\frac{\pi}{2}|\tau|+\ln|\sigma(z)|.\nonumber
\end{eqnarray}
{}From these equations it follows that
\begin{equation}
a=\frac {\pi}{2|\tau|}, \;\; {\rm and } \;\;\; b=0. \end{equation} The
constant $c$ is
determined from condition (\ref{e8}). Thus \begin{equation}
\sigma (z)=e^{\frac {\pi z^2}{2|\tau|}}\frac {\vartheta _1(z\mid \tau)
}{\vartheta'_1(0\mid \tau)}
\end{equation}
and
\begin{equation}
\tilde{G} (x)=-\frac {1}{2\pi}\ln\left|\frac {e^{\frac {\pi z^2}{2|\tau
|}}\vartheta _1(z\mid
\tau )}{\vartheta'_1(0\mid \tau )} \right|+\tilde{C}, \label{e13} \end{equation}
where the constant $\tilde{C}$ can be found from the condition \[
\int_{0}^{L_1}\int_{0}^{L_2}G_0(x)dx_1dx_2=0, \] or we can choose the other
possibility
\begin{eqnarray}
\int _{0}^{L_1}G_0(x_1,0)dx_1&=&\frac {L_1}{4\pi^2|\tau|}\sum_{n_1,
n_2}\nolimits'
\frac {\int _{0}^{1}e^{2\pi in_1x_1}dx_1}{n_1^2+|\tau|^{-2}n_2^2} \nonumber \\
&=&\frac {L_1|\tau|}{4\pi ^2}\left(\sum _{n_2=1}^{\infty}\frac {1}{n_2^2}+ \sum
_{n_2=-\infty}^{-1}\frac {1}{n_2^2}\right) \label{e14}\\ &=&\frac
{L_1|\tau|}{2\pi
^2}\sum _{n=1}^{\infty} \frac {1}{n^2}= \frac {L_1|\tau|}{12}.\nonumber
\end{eqnarray}
{}From (\ref{e.3}) and (\ref{e13}) we have \begin{equation}
G_0(x)=-\frac {1}{2\pi}\ln\left|\frac {\vartheta _1(z\mid \tau)}
{\vartheta'_1(0\mid
\tau)}\right|+\frac {x_2^2}{2L_1L_2}+\tilde{C}. \label{e15}\end{equation}
Thus from
(\ref{e14}) and (\ref{e15}) we will obtain \begin{equation} \tilde{C}=\frac
{1}{2\pi}\int
_{0}^{1}dx \ln\left|\frac {\vartheta_1(x\mid \tau)}{\vartheta'_1(0\mid
\tau)}\right|+\frac{|\tau|}{12}. \end{equation} {}From \cite{Batem}
(13.19(17)) we have
\[\ln\left|\frac {\vartheta _1(x\mid \tau)}{\vartheta'_1(0\mid \tau)}
\right|=-\ln\pi +\ln\sin
\pi x+4\sum _{m=1}^{\infty}\frac {q^{2m}}{1-q^{2m}} \frac {1-\cos 2\pi
mx}{2m},\]
where $q=e^{i\pi \tau}$.

Then
\begin{equation}
\int _{0}^{1}dx \ln \left|\frac {\vartheta _1(x\mid
\tau)}{\vartheta'_1(0\mid \tau)}\right|=-
\ln\pi -\ln2-\ln q_0^2(\tau ), \end{equation} where $q_0^2(\tau
)=\prod_{n=1}^{\infty}(1-q^{2n})$. \par So
\begin{equation}
\tilde{C}=\frac {|\tau|}{12}- \frac {1}{2\pi}\ln(2\pi q_0^2(\tau)).
\end{equation}
Finally we get
\begin{equation}
G_0(x)=-\frac {1}{2\pi}\ln \left|\frac {\vartheta _1(z\mid
\tau)}{\vartheta' _1(0\mid
\tau)}\right|+\frac {x_2^2}{2L_1L_2}+\frac {|\tau|}{12}-\frac {1}{2\pi}\ln(2\pi
q_0^2(\tau)).
\end{equation}

\vspace{1cm}
\set
\appendix {\bf Appendix F. The propagator of fermions in the background toron
field.} \vspace{0.7cm}
\renewcommand{\theequation}{F.\arabic{equation}}

The fermion propagator $S_t(x)$
in the background toron field is defined as a solution of the equation
\begin{equation}
D_tS_t(x)\equiv \gamma_{\mu}(\partial_{\mu}-iet_{\mu})S_t(x)=-\delta (x)
\label{f1}
\end{equation}
and is periodic in the Euclidean space-time \begin{equation} S_t(x_1+ k
_1L_1,x_2+ k _2L_2)=S_t(x),\label{f2} \end{equation} where $ k _{\mu}(\mu
=1,2)$
are some integers , and is quasiperiodic in $t$-space
\begin{equation}
S_{t_{\mu}+\frac {2\pi}{eL_{\mu}}m_{\mu}}(x)=e^{\frac {2\pi i}{L_1} m_1x_1+\frac
{2\pi i}{L_2}m_2x_2}S_t(x),\label{f3} \end{equation} where $m_{\mu}(\mu
=1,2)$ are
some integers, too. Note that $[S_t(x)]=[l]^{-1} $.
One can easily find that the solution of (\ref{f1}), which obeys the
conditions (\ref{f2})
and (\ref{f3}) has the following form \begin{equation} S_t(x)=-
D_tG_t(x)=\gamma_\mu S_t^{(\mu)}(x),\label{f4} \end{equation} where
\begin{equation}
G_t(x)=\frac {1}{L_1L_2}\sum _{n_i}\frac {e^{\frac {2\pi i}{L_1}(n_1x_1+
|\tau|^{-
1}n_2x_2)}}{\left (\frac {2\pi}{L_1}\right )^2[(n_1-\tilde
{t}_1)^2+|\tau|^{-2}(n_2-
\tilde{t}_2)]}. \end{equation}
It can be shown (see \cite{Weil}) that this function has a following
representation
\begin{eqnarray}
G_t (x)&=&\frac {1}{4\pi}{\sum_{\mu\nu}}' \frac {e^{-2\pi \mid (-\frac {x_2}
{L_2}+\nu)(\mu -\tilde{t}_1)\mid|\tau|}}{|\mu -\tilde{t}_1|} e^{2\pi i(\mu \frac
{x_1}{L_1}+\tilde{t}_2(\frac {x_2}{L_2}-\nu)}\nonumber \\ &&+\frac {|\tau|}{4\pi
^2}S_0\left(-\tilde{t}_2,-\frac {x_2}{L_2},1\right)e^{2 \pi i\frac
{x_1}{L_1}\tilde{t}_1},\;\;\;\;\;\; {\rm if}\;\;\;\;\; \tilde{t}_1 \in Z \\
&&+\frac {1}{4\pi^2}
S_0\left(-\tilde{t}_1, -\frac {x_1}{L_1},\frac {1}{2}\right),\;\;\; \;\;\;
{\rm if}\;\;\;\;\; \frac {x_2}{L_2}\in Z, \nonumber \end{eqnarray} where
$\tilde{t}_i =\frac
{eL_it_i}{2\pi}(i=1,2)$, prime in the sum means that the
terms with $\mu =\tilde{t}_1,$ if $\tilde{t}_1 \in Z$ and $\nu =\frac {x_2}
{L_2}$, if $
\frac {x_2}{L_2}\in Z$ should be omitted. \begin{equation}
S_0\left(-\tilde{t}_2,-\frac
{x_2}{L_2},1\right)={\sum_{\mu}}^{\ast} \frac {e^{2\pi i \frac
{x_2}{L_2}\mu}}{(\tilde{t}_2-\mu )^2} \end{equation} \begin{equation}
S_0\left(-\tilde{t}_1,-\frac {x_1}{L_1},\frac
{1}{2}\right)={\sum_{\mu}}^{\ast} |-\tilde{t}_1
+\mu|^{-1}e^{2\pi i\frac {x_1}{l_1}\mu}
\end{equation}
for $0\leq\tilde{t}_1<1$, and
\begin{eqnarray}
S_0\left(-\tilde{t}_1 -\frac {x_1}{L_1}, \frac {1}{2}\right)
&=&\ln\left|2\sin \frac {\pi
x_1}{L_1}\right|^2+\frac {1}{\tilde {t}_1}\nonumber \\ &+&\tilde{t}_1\sum _{\mu
=1}^{\infty}\left(\frac {e^{2\pi i \frac {x_1}{L_1}\mu} }{\mu (\mu -
\tilde{t}_1)}+ \frac {e^{-
2\pi i \frac {x_1}{L_1}\mu }}{\mu (\mu + \tilde{t}_1)}\right) \end{eqnarray}
for $0<\tilde{t}_1<1$. Star in the sum means that if $\tilde{t}_2
(\tilde{t}_1)$ is an
integer the term with $\mu = \tilde{t}_2 (\mu=\tilde{t}_1) $ should be
omitted. \par
One can prove that the $S_t^{(\mu )}(x)$ functions defined in (\ref{f4})
have the
following short distance behaviour \begin{equation} S_t^{(\mu )}(x)=\frac
{e^{iet_{\mu}x_{\mu}}}{2\pi}\left(\frac{x_{\mu}} {|{\bf x}|^2}+K_{\mu}(t)\right
)+O(|x|),\label{f10} \end{equation} where
\begin{equation}
K _1=-iet_1+\frac {1}{2L_1}\left (\frac {\vartheta
'_1(\bar{t})}{\vartheta_1(\bar{t})} -\frac
{\vartheta '_1(t)}{\vartheta_1(t)} \right ),\label{f11} \end{equation}
\begin{equation}
K _2=\frac {i}{2L_1}\left (\frac {\vartheta '_1(\bar{t})}
{\vartheta_1(\bar{t})}+\frac{\vartheta '_1(t)}{\vartheta_1(t)} \right
)\label{f12}
\end{equation}
and $t=\tilde{t}_2 +i|\tau|\tilde{t}_1,\;\;\; \bar {t}=\tilde{t}_2-i|
\tau|\tilde{t}_1.$
\par
It can easily be  checked that the functions which obey the periodicity
conditions
(\ref{f2}), (\ref{f3}) and short distance behaviour (\ref{f10}) have the
following form
\begin{equation}
S_t ^{(1)}(x)=\frac {1}{4\pi L_1}\left\{e^{ex_2t_+}\frac{\vartheta'_1(0)}
{\vartheta_1(\bar{t})}\frac {\vartheta_1(z+\bar{t})}{\vartheta _1(z)}+e^{-ex_2t_-}\;
\frac{\vartheta'_1(0)}{\vartheta _1(t)}\frac {\vartheta _1(t-\bar{z})}
{\vartheta
_1(z)}\right\} \end{equation} \begin{equation} S_t ^{(2)}(x)=\frac {1}{4\pi
iL_1}\left\{e^{-ex_2t_-}\frac{\vartheta'_1(0)}{\vartheta_1 (t)}\frac
{\vartheta _1(t-
\bar{z})}{\vartheta _1(\bar{z})}- e^{ex_2t_+}\;\frac{\vartheta'_1(0)}
{\vartheta_1(\bar{t})}\frac {\vartheta _1(z+\bar{t})}{\vartheta _1(z)}\right\},
\end{equation} where $t_{\pm}= t_1\pm it_2$, or in the matrix form
\begin{equation}
S_t(x)=\frac {e^{iet_{\mu }x_{\mu}}}{2\pi L_1}\left ( \begin{array}{cc} 0&\frac
{\vartheta'_1(0)\vartheta _1(z+\bar{t})}{\vartheta
_1(\bar{t})\vartheta_1(z)}e^{-
ieL_1t_1z} \\ -\frac {\vartheta'_1(0)\vartheta _1(\bar{z}-t)}{\vartheta
_1(t)\vartheta_1(\bar{z})}e^{-ieL_1t_1 \bar{z}}&0\\ \end{array} \right).
\end{equation}

Note that $S_t(x)$ becomes singular for $t=0$. This singularity is caused by the
constant solution of the Dirac equation with $t=0$, which represents zero
modes in
the trivial sector. In the path integral it is compensated by the zero of
the Boltzmann
factor of the induced toron action $Z_{tor}(t)$.

\vspace{1cm}
\set
\appendix{\bf Appendix G. The propagator of fermions in the external field
from the
topologically nontrivial class ($k\neq0$).} \vspace{0.7cm}
\renewcommand{\theequation}{G.\arabic{equation}}
\newcommand{\gn}[1]{\renewcommand{\theequation}{G.\arabic{equation}#1}}
\newcommand{\fgn}{\renewcommand{\theequation}{G.\arabic{equation}}}

Fermion propagator in the background field $A_\mu^{(k)}$ \begin{eqnarray}
S_{\alpha\beta}^{(k)}(x,y;A)=\langle\psi_\alpha(x)\bar{\psi}_\beta(y)
\rangle_\psi^{(k)}
\label{g1}
\end{eqnarray}
is a solution of the equation
\begin{eqnarray}
DS^{(k)}(x,y;A)=-\delta^{(2)}(x-y)+{\cal P}_0(x,y), \end{eqnarray} where ${\cal
P}_0(x,y)$ is the projector operator onto the space of the zero modes.

Using (\ref{a17})-(\ref{a19}) the Dirac operator $D$ can be written as follows
\begin{eqnarray}
D=e^{i\beta_+(x)}\gamma_{\mu}\partial_{\mu}e^{-i\beta_-(x)} \end{eqnarray} with
\begin{eqnarray}
\beta_{\pm}(x)=e\{a(x)+t_{\mu}x_{\mu}\pm i\gamma_5[b(x)-\frac{\pi kx^2}
{2eL_1L_2}]\}.
\end{eqnarray}
Then the propagator $S^{(k)}(x,y)$ can be expressed in the following form
\begin{eqnarray}
S^{(k)}(x,y;A)=e^{i\beta_-(x)}Q^{(k)}(x,y)e^{-i\beta_+(y)}\label{g5}
\end{eqnarray}
and $Q^{(k)}$obeys the equation
\begin{eqnarray}
\gamma_{\mu}\partial_{\mu}Q^{(k)}(x,y)=-\delta^{(2)}(x-y)+\tilde {{\cal
P}}_0(x,y),\label{g6}
\end{eqnarray}
where
\begin{eqnarray}
\tilde{{\cal P}}_0(x,y)=e^{-i\beta_+(x)}{\cal P}_0(x,y)e^{i\beta_+(y)}.
\end{eqnarray}

Further, we consider in detail only the case $k>0$, because for $k<0$
everything is
similar. For the trasformed projection operator we have \begin{eqnarray}
\tilde{\cal
P}_0(x,y)=\sum_{n=0}^{k-1}\tilde{\chi}^{(n)}(x)\bar{\tilde{\chi}}^{(n)}(y),
\label{g8}
\end{eqnarray}
where $\tilde{\chi}^{(n)}$ are transformed zero modes with positive chirality.

One can introduce a new spinor $\eta^{(n)}(x)$ such that

\begin{eqnarray}
\gamma_{\mu}\partial_{\mu}\eta^{(n)}(x)=\tilde{\chi}^{(n)}(x). \label{g9}
\end{eqnarray}
Then we define a new function

\begin{eqnarray}
\tilde{Q}^{(k)}(x,y)=Q^{(k)}(x,y)+\sum_{n=0}^{k-1}\eta^{(n)}(x)\bar{\tilde
{\chi}}^{(n)}(y)
\label{g10}
\end{eqnarray}
and from (\ref{g6}),(\ref{g8}) and (\ref{g9}) obtain for it

\begin{eqnarray}
\gamma_{\mu}\partial_{\mu}\tilde{Q}^{(k)}(x,y)=-\delta^{(2)}(x-y).
\end{eqnarray}
Since $\{D,\gamma_5\}=0,\gamma_5S^{(k)}\gamma_5=-S^{(k)}$ and
$\gamma_5\tilde{Q}^{(k)} \gamma_5=-\tilde{Q}^{(k)}$ the matrix
$\tilde{Q}^{(k)}$ has
only nonzero off-diagonal elements
\begin{eqnarray}
\tilde{Q}^{(k)}=\left( \begin{array}{c}0\;\;\;
\tilde{Q}_{12}^{(k)}\\\tilde{Q}_{21}^{(k)}\;\;\; 0
\end{array} \right). \end{eqnarray}
In terms of complex variables $z=\frac{x_1+ix_2}{L_1}$ and
$w=\frac{y_1+iy_2}{L_1}$ this equation can be rewritten as follows
\begin{eqnarray}
2L_1\partial_z\tilde{Q}_{21}^{(k)}(z,w)=-\delta^{(2)}(z-w), \end{eqnarray}
\begin{eqnarray}
2L_1\partial_{\bar{z}}\tilde{Q}_{12}^{(k)}(z,w)=-\delta^{(2)}(z-w). \label{g14}
\end{eqnarray}

Let us consider the function $\tilde{Q}_{12}^{(k)}(z,w)$ (for
$\tilde{Q}_{21}^{(k)}(z,w)$
everything can be done in the same way). {}From the definition of the fermion
propagator (\ref{g1}), from relations (\ref{g5}), (\ref{g10}) and from the
periodicity
conditions (1.2) for Fermi fields, which in this particular case have the
form given in
(1.22), the periodicity conditions on $\tilde{Q}_{12}^{(k)}(z.w)$ can be
found to be
\gn{a}
\begin{eqnarray}
\tilde{Q}_{12}^{(k)}(z+1,\bar{z}+1;w,\bar{w})= e^{-iet_1L_1+ \frac{\pi
k}{|\tau|}z+\frac{\pi k}{2|\tau|}}
\tilde{Q}_{12}^{(k)}(z,\bar{z};w,\bar{w}), \end{eqnarray} \add
\gn{b}
\begin{eqnarray}
\tilde{Q}_{12}^{(k)}(z+\tau ,\bar{z}+\bar{\tau};w,\bar{w})= e^{-iet_2L_2 - i\pi
kz+\frac{\pi k}{2}|\tau|} \tilde{Q}_{12}^{(k)}(z,\bar{z};w,\bar{w}),
\end{eqnarray}
\add
\gn{c}
\begin{eqnarray}
\tilde{Q}_{12}^{(k)}(z,\bar{z};w+1,\bar{w}+1)= e^{iet_1L_1- \frac{\pi
k}{|\tau|}z-\frac{\pi
k}{2|\tau|}} \tilde{Q}_{12}^{(k)}(z,\bar{z};w,\bar{w}), \end{eqnarray} \add
\gn{d}
\begin{eqnarray}
\tilde{Q}_{12}^{(k)}(z,\bar{z};w+\tau ,\bar{w}+\bar{\tau})= e^{-iet_2L_2+
\pi kw-
\frac{\pi k}{2}|\tau|}
\tilde{Q}_{12}^{(k)}(z,\bar{z};w,\bar{w}).\label{g15} \end{eqnarray} \fgn
The general form of $\tilde{Q}_{12}^{(k)}(z,w)$ that satisfies (\ref{g14}) is
\begin{eqnarray}
\tilde{Q}_{12}^{(k)}(z,w)=\frac{1}{2\pi L_1}\frac{\vartheta_1^{\prime}(0)}
{\vartheta_1(z-w)}q(z,w),
\end{eqnarray}
where $q(z,w)$ should be chosen to obey the periodicity conditions in
(G.15) , to
have no poles for $z-w$ and to be 1 at $z=w$. A solution that satisfies these
conditions is \begin{eqnarray}
\tilde{Q}_{12}^{(k)}(z,w)= \frac{1}{2\pi L_1}\frac{\vartheta_1^{\prime}(0)}
{\vartheta_1(z-w)}\frac{\vartheta_1(z-w+\bar{t})}{\vartheta_1(\bar{t})}
\frac{\vartheta_3(kz|k\tau )}{\vartheta_3(kw|k\tau )}\nonumber\\ \times e^{-
ieL_1t_1(z-w)+\frac{\pi k}{2|\tau |}(z^2-w^2)}. \end{eqnarray}

For $\tilde{Q}_{21}^{(k)}(z,w)$ one obtains the similar expression
\begin{eqnarray}
\tilde{Q}_{21}^{(k)}(z,w)= -\frac{1}{2\pi L_1}\frac{\vartheta_1^{\prime}(0)}
{\vartheta_1(\bar{z}-\bar{w})}\frac{\vartheta_1(\bar{z}-\bar{w}+t)}
{\vartheta_1(t)}
\frac{\vartheta_3(k\bar{z}|k\tau )}{\vartheta_3(k\bar{w}|k\tau
)}\nonumber\\ \times e^{-
ieL_1t_1(\bar{z}-\bar{w})+\frac{\pi k}{2|\tau |}(\bar{z}^2-\bar{w}^2)}.
\end{eqnarray}


\begin{thebibliography}{99}

\bibitem{Schwi}
J. Schwinger, Phys. Rev. {\bf 128} (1962) 2425. \bibitem{Low} J.H.
Lowenstein and
J.A. Swieca, Ann. Phys. (N.Y.) {\bf 68} (1971) 172. \bibitem{NS}
N.K. Nielsen and B. Schroer, Nucl. Phys. {\bf B120} (1977) 62; K.D. Rothe
and J.A.
Swieca, Ann. Phys. {\bf 117} (1979) 382; M. Hortascu, K.D. Rothe and B. Schroer,
Phys. Rev. {\bf D20} (1979) 3203; N.V. Krasnikov et.al, Phys. Lett. {\bf
B97} (1980)
103; R.Roskies and F. Schaposnik, Phys. Rev. {\bf D23} (1981) 558. \bibitem{KJ}
K.Johnson, Phys. Lett. {\bf 59} (1963) 253; R. Jackiw in: {\it Relativity,
Groups and
Topology II, Les Houches 1983} edited by B.S. De Witt and R. Stora, North
Holland
1984.
\bibitem{CKS}
A. Casher, J. Kogut, L. Susskind,
Phys. Rev. {\bf D10} (1974) 732; S. Coleman, R. Jackiw, L. Susskind, Ann.
Phys. {\bf
93} (1975) 267.
\bibitem{RW}
A.K. Raina and G. Wanders, Ann.
Phys. {\bf 132} (1981) 404; \\A.Z. Capri and R. Ferrari, Nuovo Cimento {\bf A62}
(1981) 273; \\P. Becher, Ann. Phys. {\bf 146} (1983) 223. \bibitem{Camil} C.
Jayewardena, Helv. Phys. Acta {\bf 61} (1988) 636. \bibitem{Man} N.S. Manton,
Ann. Phys. (N.Y.) {\bf 159} (1985) 220. \\ J.E. Hetrick and Y. Hosotany,
Phys. Rev. {\bf D38} (1988) 2621. \\ S. Iso and H. Murayama, Progr. Theor.
Phys. {\bf
84} (1990) 142. \bibitem{Aza}
S.I.Azakov, The Schwinger Model on the Torus. Proceedings of the Course and the
Conference on Path Integration. Trieste 26 August-10 September. 1991,
pp.543-548. \bibitem {JO1}
H.Joos, Helv.Phys.Acta {\bf 63} (1990) 670, Nucl.Phys.(Proc.Suppl.){\bf
B17} (1990)
704. \bibitem {JA}
H.Joos and S.I.Azakov, Helv.Phys.Acta {\bf 67} (1994) 723. \bibitem {SW}
I.Sachs
andA.Wipf, Helv.Phys.Acta {\bf 65} (1992) 652. \bibitem {Kao} Y.C.Kao, preprint
UCB-PTH 83/14, 1983.
\bibitem {Gil}
P.B. Gilkey, {\it Invariance Theory, The Heat Equation and the
Atiyah-Singer Index
Theorem}, Publish and Perish Inc. 1984. \bibitem {Pat} M.F.Atiyah,
W.K.Patodi and
I.M.Singer,
Math.Proc.Camb.Philos. Soc.,{\bf 79} (1978) 71 \bibitem {Mum} D.Mumford,
{\it Tata
Lectures on Theta}, Birkh\"{a}user Verlag. 1983. \bibitem {LILEU}
G.'tHooft,Phys.Rev.{\bf D14} (1976) 3432. \\ H.Leutwyler, Phys. Lett.{\bf 152B}
(1985) 65. \bibitem {Whit}
E.T. Whittaker and G.N.Watson, {\it Modern Analysis}, p.487, Cambridge
University
Press, 1969. \bibitem {Batem}
A.Erdelyi (Director), {\it Higher Trancedental Functions}, Vol.2 Chapter
13, Vol.3,
Chapter 17, 1953.
\bibitem {Abram}
M.Abramowitz and I.A. Stegun, {\it Handbook of Mathematical Functions}, Dover,
New York, 1970.
\bibitem {Weil} A. Weil, {\it Elliptic Functions
according to Eisenstein and Kronecker}, Springer-Verlag, 1976.
\bibitem{Tolke}
F.T\"{o}lke, {\it Praktische Funktionenlehre}, Zweiter Band,
Springer-Verlag, 1966.
\bibitem {Fay}
A.Fayyazuddin, T.H.Hansson, M.A.Nowak, J.J.M. Verbaarschot and I. Zahed,
Nucl.Phys.{\bf B425} (1994) 553.

\end{thebibliography}
\end{document}